\documentclass[journal=nalefd,manuscript=letter]{achemso}
\usepackage[version=3]{mhchem} 
\usepackage{graphicx}
\usepackage{dcolumn}
\usepackage{bm}
\usepackage{comment}
\usepackage{epstopdf}
\usepackage{amsmath}
\usepackage{hyperref}
\author{I.~G. Rebollo$^{\S}$, F.~C. Rodrigues-Machado$^{\S}$, W. Wright, G.~J. Melin, and A.~R. Champagne}
\email{a.champagne@concordia.ca}
\affiliation{$^{\S}$These two authors made equal contributions. \linebreak Department of Physics, Concordia University, Montr\'{e}al, Qu\'{e}bec, H4B 1R6, Canada}
\title[\texttt{achemso}]
{Thin-suspended 2D Materials: Facile, Versatile, and Deterministic Transfer Assembly}
\begin{document}

\begin{abstract}
We report a deterministic 2D material (2DM) transfer method to assemble any-stacking-order heterostructures incorporating suspended ultra-thin 2D materials, such as single-layer graphene (SLG) and bilayer graphene (BLG). The transfer procedure relies on a single-step preparation nitrocellulose micro-stamp, which combines both outstanding adhesion and softness. It permits the dry pick-up of naked 2D crystals (graphene, MoS$_2$, and hBN) directly from a SiO$_2$ substrate, and to precisely transfer them on substrates or trenches. Optical and Raman data show that no significant defect is introduced upon transfer, even in suspended SLG and BLG. The areas transferred range up $\sim$ 1000 $\mu$m$^2$ on substrate. High-yield transfer of suspended ultra-thin 2DM does not require critical point drying for areas up to 15 $\mu$m$^2$ or suspension heights down to 160 nm. To demonstrate the method's capabilities, we assembled on-substrate and suspended optical cavities tuning BLG's Raman scattering intensity by factors of 19 and 4, respectively. This resilient and rapid 2DM transfer procedure will accelerate the fabrication of many heterostructures and permit versatile suspension of 2DMs for research in twistronics, straintronics, and nano-opto-electro-mechanical systems (NOEMS).
\end{abstract}

Advances in the ability to mechanically transfer, align, and stack 2D materials (2DMs)\cite{Frisenda18, Fan20}, to form pristine heterostructures\cite{Novoselov16,Liu19_2}, have greatly accelerated experiments in quantum electron transport \cite{Shen20, Li19_3, Yankowitz19, Song18} and optoelectronics \cite{Rivera18, Deilmann20} over the last several years. The wide array of 2DM transfer methods now available \cite{Fan20, Frisenda18} can assemble most of the possible vertically-stacked heterostructures, with major exceptions for suspended 2DMs. Moreover, the state-of-the-art 2D transfer methods \cite{Wang13,Frisenda18, Fan20} are complex and time consuming since they often requires using a stamp made with two separate films, such as polypropylene carbonate (PPC) and polydimethylsiloxane (PDMS). There is a need to develop a single-material stamp combining the adhesion of PPC and softness of PDMS for more resilient, versatile, and faster transfers.

Presently, there is no flexible (any 2DM, any stacking order) and deterministic (with alignment) procedure to assemble heterostructures incorporating 2DMs and layers of vacuum/air and suspended ultra-thin 2DMs. Indeed, previous transfer methods for heterostructures incorporating \textit{suspended} 2DMs have either used thick-suspended crystals\cite{Castellanos14, Singh14, Wakafuji20}, non-deterministic transfer of thin-suspended 2DMs\cite{Li15}, or developed single-purpose custom micro/nanofabrication routes for each device geometry\cite{Liu19_1, Zhang19_2, Yigen14, Bolotin08}. There are many motivations for integrating suspended 2DMs in precisely assembled heterostructures. Such devices would permit unprecedented levels of simultaneous control of electronics, mechanics, optics, and their interactions in nano-opto-electro-mechanical systems (NOEMS)\cite{Singh14,Midolo18,Roy18}. For example, vacuum layers offer a uniquely different index of refraction to optimize exciton binding energy and lifetime in 2DMs \cite{Florian18}. In quantum transport studies of twisted bilayers (twistronics)\cite{Kim16, Hu2020, Weston20}, properly designed suspension would decouple the mechanically sensitive bilayers from the substrate to permit strain-engineering\cite{Naumis17,McRae19, Zhang20} of their quantum phases \cite{Cao18,Xian19,Li19_3}. Additionally, stacking 2D suspended nano-electro-mechanical systems (NEMS) on top of 2D mirrors can create optical cavities\cite{Song15, Casalino17, Epstein20} enhancing light-matter interactions and hybridizing photonics with NEMS physics \cite{Midolo18, Eggleton19, Li19_2, Zhang19_2}.

Here we present the development of a 2DM transfer method able to dry pick-up ultra-thin naked (i.e. not encapsulated) crystals directly from SiO$_2$, and then transfer them on substrates or trenches using only microliters of mild solvents. Our facile stamping procedure relies on a nitrocellulose micro-stamp ($\sim$ 200 $\mu$m wide) deposited on a glass slide actuated by a micromanipulator. This microstamp is much simpler to prepare than previously used PPC/PDMS stamps \cite{Wang13}. The micro-stamp improves both the optical contrast and manipulation of thin 2DMs. It takes less than one hour to complete the transfer of a crystal (graphene, MoS$_2$, and hBN) with areas up to $\sim$ 1000 $\mu$m$^2$. A single robust protocol was developed to transfer and align crystals onto different substrates (SiO$_2$, hBN, aluminum). We verify via optical imaging and Raman spectroscopy, that the crystals are not damaged during transfer and can be aligned with a $\sim$ 1 $\mu$m accuracy. We demonstrate the assembly of high-quality heterostructures both on substrate (e.g. BLG/hBN/Al) and suspended over hBN trenches (e.g. BLG/air/SiO$_2$/Si). No critical point drying is required to transfer ultra-thin 2D areas up to 15 $\mu$m$^2$ or suspension heights down to 160 nm. A first application of this transfer assembly is demonstrated by fabricating optical cavities able to engineer the Raman scattering intensity (Raman factor) of BLG. We find a quantitative agreement between first principle calculations and experimental data of the BLG Raman factor, $F_{BLG}$, and for its underlying exclusive light absorption $A_{BLG}$, on several BLG/hBN/Al and BLG/Air/SiO$_2$/Si cavities. The $F_{BLG}$ is tuned by factors of 19 and 3.8 in supported and suspended BLG heterostructures, respectively. Given the unique electronic and optical properties of BLG, maximizing $F_{BLG}$ and $A_{BLG}$ would bring new opportunities in light harvesting and photo-electric devices \cite{Song15,Casalino17,Deilmann20}.

\begin{figure*}
\includegraphics[width=6.25in]{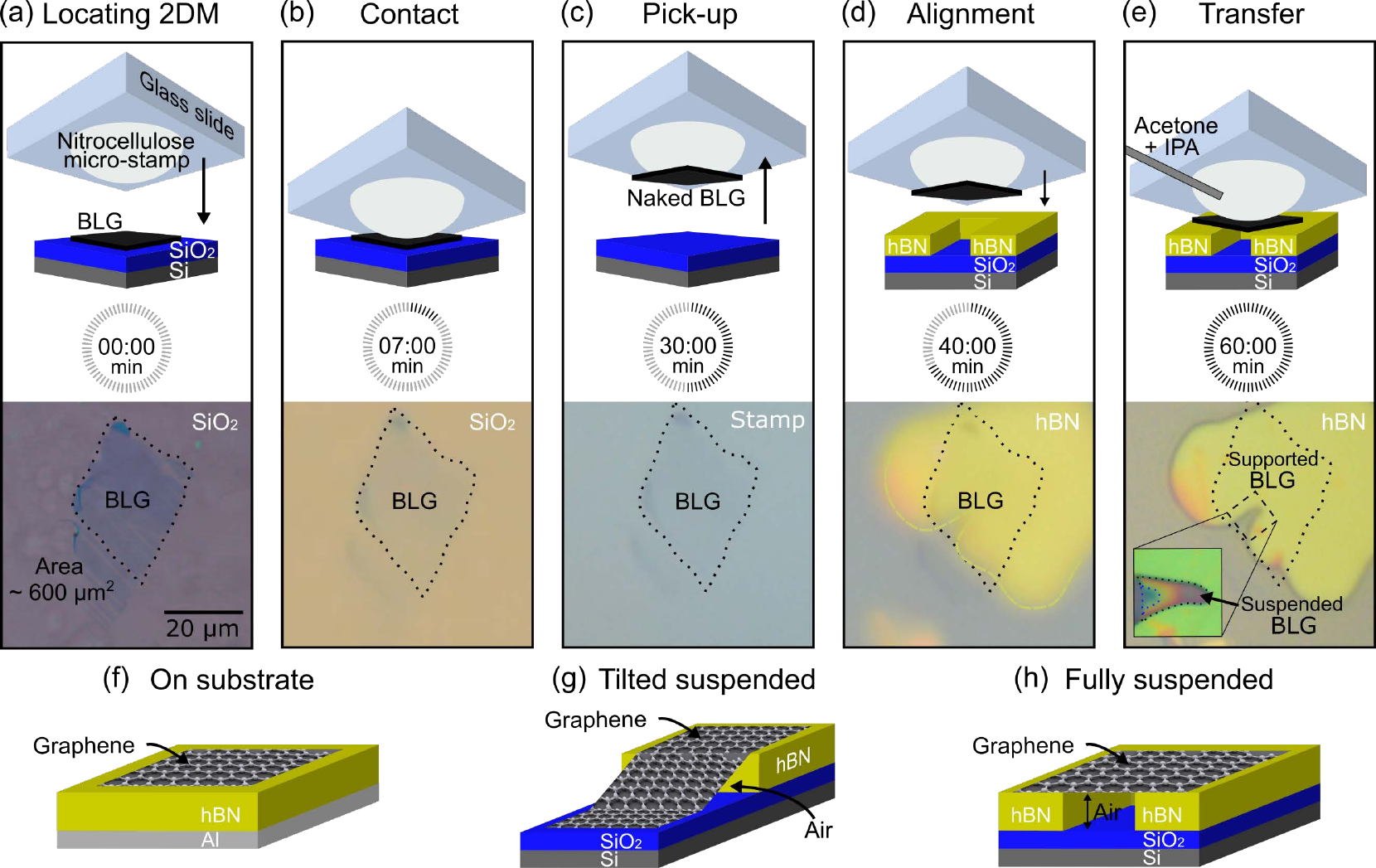}
\caption{Step-by-step deterministic assembly of heterostructures incorporating thin-suspended 2D crystals. (a) Locating the 2DM to be transferred through the transparent nitrocellulose micro-stamp. (b) Pressing the micro-stamp directly on a 2D crystal to promote adhesion. (c) Picking-up the 2D crystal directly from a SiO$_2$ substrate or another substrate. (d) Micron-precision alignment of the naked (not encapsulated) 2D crystal above a new substrate. (e) Transferring the 2DM by lowering, pressing, and dissolving the nitrocellulose stamp with droplets of mild solvents. (f)-(h) Three types of transferred heterostructures acting as optical cavities: on-substrate, tilted-suspended, and fully-suspended.}
\end{figure*}

Deterministic stamping of any-stacking-order and suspended ultra-thin 2DMs such as SLG and BLG has been out of reach so far. A first challenge explaining this is that while SiO$_2$ substrates permit an easy identification of thin exfoliated crystals \cite{Blake07}, dry pick-up of naked 2DMs on SiO$_2$ has not generally been possible due to the strong substrate adhesion \cite{Fan20}. Secondly, the controlled pick-up and transfer of one crystal, leaving nearby flakes untouched, is not trivial and best done with a three-dimensional micro-stamp \cite{Wakafuji20}. Lastly, the stamping process of thin-suspended crystals requires finesse to avoid tearing of the crystal, or collapse of the suspended layer due to capillary forces \cite{Ma17}. Figure 1 summarizes the main steps of our deterministic any-stacking-order and thin-suspended 2DM transfer. It rests on a dry nitrocellulose micro-stamp which strongly adheres to 2DMs for pick-up. This stamp acts as a micro-lens improving the imaging contrast of ultra-thin crystals, and as a micro-manipulator able to gently handle them. The micro-stamp can be dissolved with mild solvents. Our stamping instrumentation is similar to the one used in previous transfer methods \cite{Castellanos14, Frisenda18} (see the Supporting Information S1). It consists of a rotating ($\theta$) stage holding the substrate, $x-y-z$ micro manipulators translating the glass slide and micro-stamp, and a long working distance optical microscope for live monitoring the transfer. A complete transfer procedure as shown in Figure 1 can be done under 60 minutes. The procedure is the same for various materials: SLG, BLG, FLG, hBN, and MoS$_2$, and does not require any temperature control \cite{Pizzocchero16}, microfabrication step, or harsh chemical.

The transfer procedure starts with Figure 1a, where we position a micro-stamp above the 2DM to be picked-up. The preliminary preparation of the micro-stamp, substrates, and 2D crystals is discussed in detail in the Supporting Information S1. A fresh micro-stamp is prepared before each transfer. The stamp is roughly a half-dome with $x-y-z$ dimensions of $\sim$ 500 $\mu$m, while its contact area during transfer is limited to $\sim$ 200 $\mu$m $\times$ 200$\mu$m. To create a micro-stamp, we deposit a sub-microliter amount of a commercially available nitrocellulose polymer solution (Extra Life$^{TM}$ No Chip Top Coat - Revlon) on a glass slide, and let it dry for approximately 7 minutes. The stamp permits high resolution optical imaging of 2DMs when looking through it, as shown for a typical BLG crystal at the bottom of Figure 1a. Images of the same crystal are shown at each stage of the transfer in Figure 1b-e. In Figure 1b, the micro-stamp is first carefully aligned with the target crystal, and then brought down at a speed of $\sim$ 50 $\mu$m/s until it makes contact (detected by a sudden change of color). The stamp is pressed down gently so that it only makes contact with the crystal and its immediate surrounding area. A one-way pressing down motion is required (i.e. no back and forth) to avoid deforming the micro-stamp and inducing folding of the crystal \cite{Wakafuji20}. The contact is maintained for $\sim$ 20 minutes to promote strong adhesion.

The stamp is then raised to pick-up the 2DM from the substrate, as shown in Figure 1c. Such a dry pick-up directly from SiO$_2$ is not possible with most previously reported stamps such as polydimethylsiloxane (PDMS) \cite{Castellanos14, Tao18}, polymethylmethacrylate (PMMA) \cite{Uwanno15}, thermal release tape (TRP) \cite{Kim15}, polyvinyl alcohol (PVA) \cite{Frisenda18}, and polypropylene carbonate (PPC) \cite{Kim16, Kinoshita19}. A key parameter to ensure a defect-free pick-up is to control the raising speed. We found that $\sim$ 250 $\mu$m/s is ideal for multilayer 2DMs, and $\sim $ 500 $\mu$m/s is best for ultra-thin materials. In Figure 1d, the target substrate is first placed underneath the micro-stamp/2DM assembly with micron precision, and then the stamp is lowered. An example of the alignment precision is seen by comparing the contour of the BLG crystal in Figures 1d and 1e, and found to be $\sim$ 2 $\mu$m in this instance. The stamp is lowered ($\sim$ 5 $\mu$m/s) until it contacts the new substrate. We monitor the pressure applied during the transfer by ensuring that the stamp does not contact the SiO$_2$ immediately surrounding the hBN substrate.

Figure 1e shows how the stamp is pressed against the target substrate. The following step is to dissolve the stamp using sub-milliliter volumes of acetone, followed by isopropyl alcohol (IPA), to cleanly transfer the 2DM (Supporting Information S1). A pipette is first used to insert acetone into the spacing between the glass slide anchoring the micro-stamp and the substrate. The acetone rapidly dissolves the nitrocellulose and releases the 2DM. Once the stamp is dissolved, and before the acetone evaporates, we raise the glass slide by $\sim$ 500 $\mu$m and use the same pipette to flush with IPA. We repeat this IPA flushing to clean any residue. At this point the 2DM has been transferred to the new substrate, but is covered with IPA (a low surface tension solvent). For suspended crystal transfers, we control the drying (evaporation rate) of IPA to avoid large capillary forces. By raising or lowering the glass slide we can modulate the evaporation rate (see Supporting Information S1). This permits a delicate transfer of suspended crystals, such as the suspended BLG region in Figure 1e. The complete transfer sequence is shown in the Supporting Movie 1. Figures 1f-h show the three main heterostructure geometries we discuss in the rest of this work: (f) on-substrate BLG/hBN/Al, (g) tilted-suspended BLG/air(variable thickness)/SiO$_2$/Si, and (h) fully-suspended BLG/air/SiO$_2$/Si.

\begin{figure}
\includegraphics[width=6.25in]{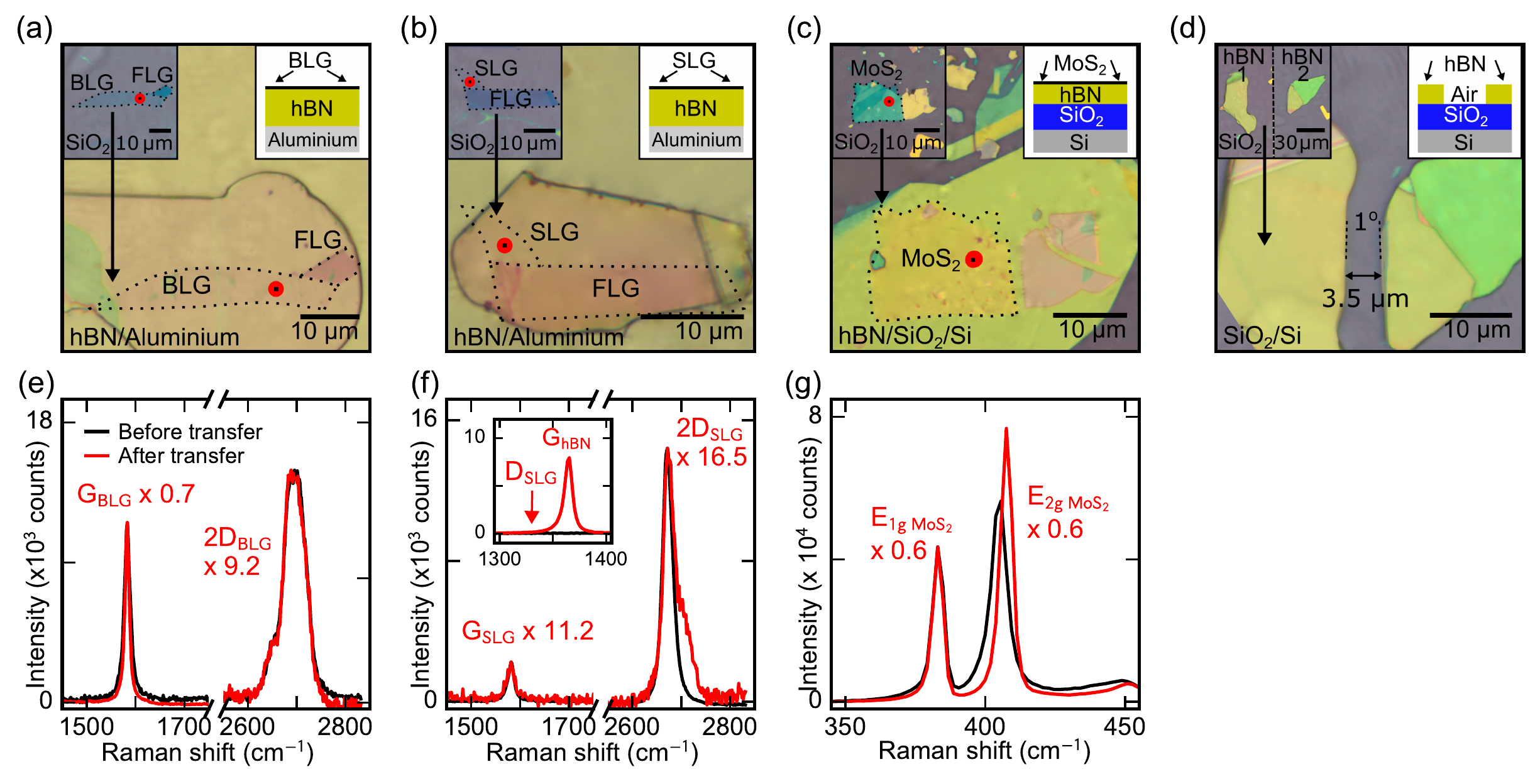}
\caption{Transfer of naked 2DMs from SiO$_2$ to another substrate. (a), (b), and (c) Optical images of large area crystals of BLG/FLG, SLG/FLG, and 10-nm-thick MoS$_2$, respectively, after their transfer. The top left insets show the crystals before pick-up on SiO$_2$ substrates. The top-right inset shows the cross-section of the final heterostructures. (d) The top-left inset shows two widely separated thick hBN crystals. These crystals were transferred one-by-one with translational and rotational alignment to create the hBN trench shown in the main panel. The top-right inset shows the cross-sectional trench geometry. (e),(f), and (g) Raman spectra taken at the red-marker locations in panels (a),(b), and (c) respectively. The black data was acquire before the crystal transfer, and the red data after the transfer.  The inset of panel (f) highlights the absence of the defect related Raman D-peak.}
\end{figure}

In Figure 2, we present large area heterostructures incorporating SLG, BLG, FLG, hBN, and MoS$_2$(10-nm-thick) crystals, without encapsulation, which were assembled as per Figure 1. To create the heterostructures in the main panels of Figures 2a,b we first evaporated a 40 nm-thick film of Aluminum on SiO$_2$/Si substrates, while for the ones in Figure 2c,d we used bare SiO$_2$/Si substrates. The Al or Si bottom layers of the heterostructures will later on act as back-plane optical mirrors. The next step of fabrication was to deterministically transfer thick ($\sim$ 400 nm) and large area hBN crystals (up to $\sim$ 1000 $\mu$m$^2$) which will act as substrates for the ultra-thin crystals, and will later define the thickness of the planar optical cavities. Then, we picked-up the thin 2D crystals shown in the insets of Figures 2a-c. They were transferred without any tearing or folding, and very few bubbles, as visible in the main panels. The top-right insets show the cross-section of the final stacks. The transferred crystals were a few hundred $\mu$m$^2$ and included SLG, BLG, few-layer graphene (FLG), and MoS$_2$.

Figure 2d displays how we can controllably position two 2DMs at deterministic relative $x-y-\theta$ positions. Two separate transfers were used to pick-up the two hBN crystals on SiO$_2$ shown in the top-left inset of Figure 2d, and to assemble them into a narrow hBN trench. The quality of both the rotational alignment (one-degree precision) and translational (one-micron precision) are clearly visible in the main panel of Figure 2d. Such trenches can be used for a lithography-free assembly of suspended heterostructures, by stamping a 2DM on top of the trench.

The Raman data shown in Figures 2e-g were taken at the red-marker locations in Figures 2a,b,c. We acquired Raman spectra at many spatial locations (Supporting Information S2), but since there is no significant spatial variation we show a single spectrum for clarity. The black data were acquired before transfer of the crystals, and the red data after the transfers. As expected, the relative heights of the G and 2D graphene Raman peaks are different before and after transfer due to optical interferences inside of the heterostructures (discussed below). The red data in Figures 2e-g are scaled as indicated. The widths of the Raman resonances are the same before and after transfer, and there is no resolvable D-peak in the graphene spectra (inset of Figure 2f). This indicates that no microscopic disorder was introduced in the crystals during stamping. Out of the 21 on-substrate heterostructure assemblies we carried out, 18 were completely successful and similar to the ones in Figure 2 (see Supporting Information S2), two were partially successful (some tearing) but produced the desired planar heterostructures, and only one was not transferred.

\begin{figure}
\includegraphics[width=6.25in]{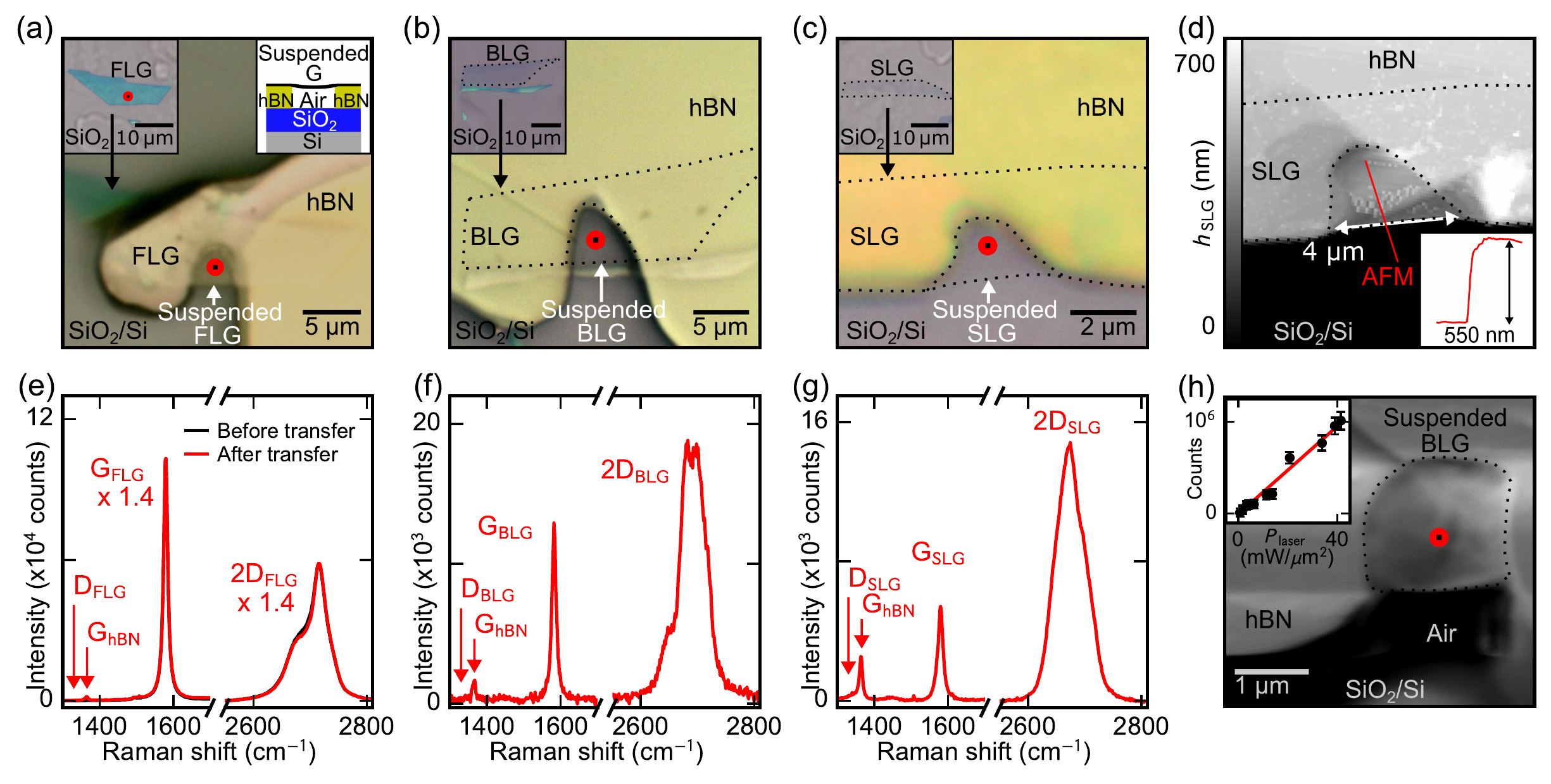}\caption{Precise stamping of thin-suspended graphene heterostructures. (a), (b), and (c) show respectively large area FLG, BLG, and SLG crystals picked-up directly from SiO$_2$ (top-left insets) and stamped on hBN trenches (main panels). The top-right inset of (a) shows the final heterostructure cross-section geometry. (d) Atomic force microscopy (AFM) image of (c). The red line marks the location of the AFM trace shown in the bottom right inset, which confirms a uniform suspension. (e), (f), and (g) Raman spectra taken at the red-marker positions in (a),(b),(c) respectively, before (black data) and after (red data) the crystal transfers. (h) Tilted scanning electron microscope (SEM) image of a suspended BLG after transfer. The inset shows the integrated Raman G-peak intensity versus laser power at the red-marker location.}
\end{figure}

Figure 3 shows four different thin-suspended-graphene/air/SiO$_2$/Si heterostructures assembled following the procedure in Figure 1. Optical images (Figures 3a-c), AFM data (Figure 3d), Raman data (Figures 3e-g), and titled-SEM imaging (Figure 3h), and detailed Supplementary Information for two additional suspended devices (Figures S8-S9), show that the thin-suspended crystals can be transferred without tearing, folding, or the introduction of microscopic defects. The transferred crystals were precisely aligned over hBN trenches, and suspended in close proximity to a back plane of SiO$_2$. The suspension height, which is crucial to maximize optical resonances or electrostatic gating, is as small as 550 nm in Figure 3, down to 340 nm in another suspended device (Supporting Information S3), and as small as 160 nm in a tilted-suspended device presented below. The 2DM suspended areas in Figure 3 range up to $\sim$ 15 $\mu$m$^2$. The U-shaped hBN trenches we used form naturally during hBN exfoliation, and are ideal lithography-free trenches for graphene suspension. Figure 3a-c, show suspended graphene crystals after transfer (before transfer in top-left insets). The final heterostructures' geometry is shown in the top-right inset of Figure 3a. The suspended FLG area's optical contrast is clearly visible in Figure 3a, and black dotted lines show the contours of the BLG and SLG crystals in Figures 3b,c. Additional evidence of the complete suspension of the SLG crystal from Figure 3c, is shown in the AFM image of Figure 3d. The bottom left inset, is an AFM trace extracted at the location of the red line in the main panel, and confirms the SLG suspension at a height of 550 nm above the substrate.

Figures 3e-g show Raman data acquired at the red-marker locations on the graphene crystals before (black data) or after (red data) suspension. Additional Raman data before and after suspension are shown in Figure S8. While the relative height of the Raman resonances changes before/after due to optical interferences inside the heterostructure, the width of the resonances remains the same and no graphene D-peak is visible after suspension. Figure 3h shows a tilted-SEM image of a suspended BLG after transfer. It demonstrates that the suspended surface is wrinkle-free and of uniform height. The top left inset shows the integrated area under the Raman G-peak (number of scattering events) as a function of laser power measured at the red marker location. The linear relationship confirms that the suspended crystal does not heat up over a laser power range which exceeds the ones we used for our Raman data acquisitions, i.e. 0 - 40 mW/$\mu$m$^2$ for 10 seconds. We carried out 16 suspended crystal transfers and 15 were successful, and similar to the ones shown in Figure 3 (see Supporting Information S3). This high-yield and precise transfer of ultra-thin suspended 2DMs could facilitate the fabrication of novel NEMS and NOEMS. We demonstrate a first example in Figure 4.

\begin{figure*}
\includegraphics [width=6.25in]{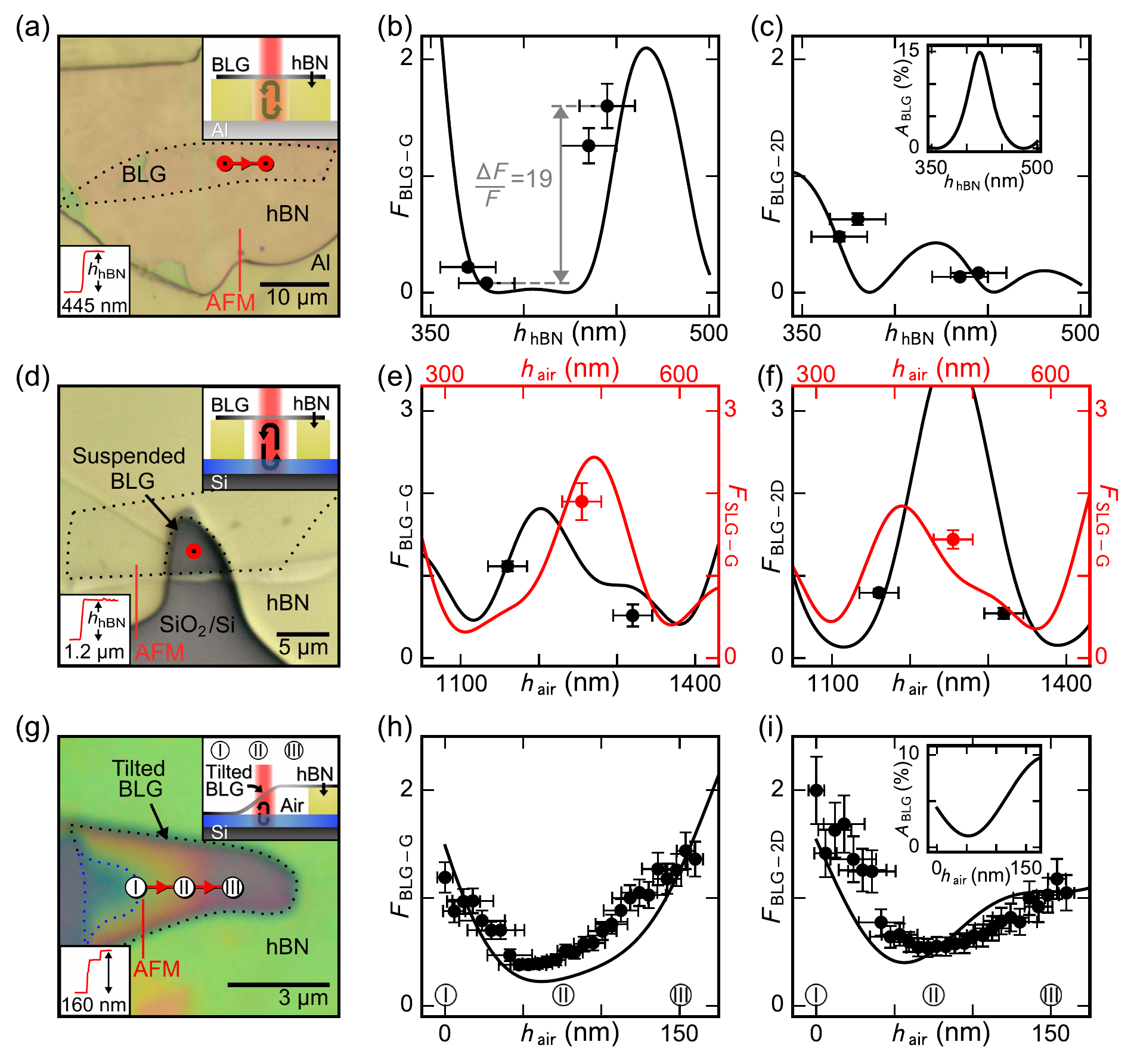}\caption{Tunable Raman factor, $F_{\mathrm{BLG}}$, and exclusive light absorption, $A_{\mathrm{BLG}}$, in BLG heterostructures. (a)  Top view of a BLG/hBN(445nm)/Al planar optical cavity. The lower-left inset shows an AFM trace taken at the red line location. (b) G-peak Raman Factor, $F_{\mathrm{BLG-G}}$, and (c) 2D-peak Raman factor, $F_{\mathrm{BLG-2D}}$, in device (a) and devices with various $h_{\mathrm{hBN}}$. The markers are the experimental data and the solid trace is zero-fit-parameter calculation. The inset of (c) shows the calculated $A_{\mathrm{BLG}}$. (d) Top view of a BLG/Air/SiO$_2$/Si device with air spacer thickness $h_{\mathrm{air}}=$ 1160 nm. (e) $F_{\mathrm{BLG-G}}$ ($F_{\mathrm{SLG-G}}$), and (f) $F_{\mathrm{BLG-2D}}$ ($F_{\mathrm{SLG-2D}}$) is shown on the left (right) axis versus $h_{\mathrm{air}}$ on the bottom (top) axis. (g) Top view of a transferred tilted-suspended BLG. In region I, the BLG sits on the SiO$_2$, while in region II it is suspended at heights ranging from 0 to 160 nm, and is held at a constant height in region III. The inset shows a cross-section of the device's geometry. (h) $F_{\mathrm{BLG-G}}$ and (i) $F_{\mathrm{BLG-2D}}$ in device (g) versus $h_{\mathrm{air}}$. The Raman data are solid markers, and the solid trace is a zero-fit-parameter calculation. The inset of (i) is the calculated $A_{\mathrm{BLG}}$ versus $h_{\mathrm{air}}$.}
\end{figure*}

Developing heterostructures for optoelectronics applications\cite{Rivera18}, and exploring 2D light-matter interactions\cite{Midolo18}, often requires increasing the very small bare light absorption in ultra-thin 2DMs \cite{Zhang19_2}. Planar heterostructures can act as optical interferometric cavities to greatly enhance light absorption and Raman scattering in 2DMs, including graphene \cite{Yoon09,Casalino17} and 2D transition metal dichalcogenides (TMDs) \cite{Epstein20}. While previous work has focused on maximizing absorption in the infrared range \cite{Casalino17, Nematpour19} and for on-substrate 2DMs, here we focus on absorption of visible light in suspended BLG. In Figure 4, we study our BLG/hBN/Al and BLG/air/SiO$_2$/Si heterostructures as optical cavities, where the BLG acts as a low-reflectivity input mirror and the substrate (Al or Si) acts as a high-reflectivity bottom mirror. We study these optical cavities using Raman spectroscopy, which provides clearly distinctive signatures of light scattering and absorption in the graphene layer than scattering in other parts of the heterostructures. The BLG Raman factor, $F_{BLG}$, is defined as the ratio of the integrated BLG Raman count in an heterostructure and the same Raman count in BLG when the crystal is entirely surrounded by vacuum, i.e. there are no optical interferences (see Supporting Information S4)\cite{Yoon09}. The quality of our structures permits a quantitative tuning of light interferences. We demonstrate experimentally a widely tunable $F_{BLG}$, and calculate the associated exclusive light absorption $A_{BLG}$. This could open up opportunities to generate strong light-matter interactions in 2D NEMS\cite{Song15}, for instance in suspended BLG twistronics\cite{Song18} NOEMS.

Figure 4a is a top view of one of four BLG/hBN/Al(40 nm) heterostructures, and its $h_{hBN}=$ 445 $\pm$ 15 nm. The red line in Figure 4a shows the location of the AFM trace displayed at the bottom left of the panel. The solid markers in Figure 4b-c show the measured G-peak and 2D-peak Raman factors, $F_{BLG-G}$ and $F_{BLG-2D}$, versus $h_{hBN}$. The solid traces are zero-fit-parameter theoretical calculations. They are based on Fresnel's equations and the known indices of refractions of the media in the heterostructures (see Supporting Information S4). To report experimental $F_{BLG}$, we first calibrated the raw experimental Raman counts with respect to the theory. This calibration was done by carefully (Supporting information S4), and reproducibly, measuring the BLG Raman counts on BLG/SiO$_{2}$(310-nm)/Si devices, for which the theoretical Raman factors are understood \cite{Yoon09}. We found that in our experimental setup, a 1.0 Raman factor corresponds to 38000 $\pm$ 1000 and 125000 $\pm$ 4000 counts of G and 2D Raman events, at a laser power density of 1.0 mW/$\mu$m$^2$ for 10 seconds. This single reference point was used to calibrate all $F_{BLG}$ values reported.

There is a good agreement between the data and model in Figures 4b-c, and $F_{BLG-G}$ can be tuned by 19 folds. This could have an application, for instance, to strongly enhance weak Raman signals predicted in many-body phase transitions \cite{Maiti17} without requiring a disruptive increase in laser power. We calculated $A_{BLG}$ as a function of $h_{hBN}$, using the same Fresnel coefficients as $F_{BLG}$, and show the result in the inset of Figure 4c. This strongly supports that the exclusive light absorption of BLG, at the 532-nm laser wavelength, can be modulated from less than 1$\%$ to over 10$\%$ in our heterostructures. Figure 4d is a uniformly suspended BLG over a hBN trench, which is itself placed on a SiO$_{2}$/Si substrate. The red line shows the location of the AFM trace shown in the bottom inset which gives $h_{hBN}$. Figure 4e-f display the measured (solid markers) and calculated (solid traces) $F_{BLG-G}$ and $F_{BLG-2D}$ versus suspension height, $h_{air}$, for the device in Figure 4d and two more (one BLG, one SLG). The left and bottom axes are for the BLG data, and the right and top axes for the SLG data.

Figure 4g shows a top view of tilted-suspended BLG/air(variable thickness)/SiO$_2$/Si heterostructure, and the top-right inset is a diagram of its cross-section. The labeled regions I, II, and III correspond to the areas of the BLG where the air-spacer thickness is respectively 0 nm, variable from 0 to 160 nm, and a constant 160 nm. A tilted structure is an ideal platform to demonstrate the tuning of light interferences as a function of $h_{air}$ in a single device. Figure 4h,i show $F_{BLG-G}$ and $F_{BLG-2D}$ data (solid markers) and model (solid traces) versus $h_{air}$ measured on this tilted device. We observe a quantitative agreement between the data and model for both Raman modes. The G Raman factor was continuously tunable by a factor up to 3.8 in Figure 4h. The horizontal error bars in Figure 4h,i arise mostly from the uncertainty on the exact suspension profile of the device in Figure 4g (see Supporting Information S4). The inset of Figure 4i presents the calculated $A_{BLG}$, which is widely tunable. We remark that $A_{BLG}$ reaches much closer to zero in the inset of Figure 4c, where the back-plane mirror used is aluminum instead of Si. While we have focused this study on lithography-free fabrication, introducing an Al mirror underneath a suspended BLG can be done, and will permit \textit{in situ} electrostatic gate-tuning of $h_{air}$, leading to vastly tunable $A_{BLG}$ and photo-electronic interactions in NOEMS\cite{Metten16}. Taken collectively, the agreement between Raman measurements and theoretical calculations in Figure 4 supports that we can transfer both suspended and on-substrate ultra-thin crystals of sufficient quality to manipulate their Raman factor, and exclusive light absorption.

In conclusion, we presented a 2DM transfer and assembly method based on a nitrocellulose micro-stamp able to stamp crystals in any-stacking order and incorporate suspended 2DMs. This method is a much needed simplification and extension of the current state-of-the-art PPC/PDMS stamping procedure \cite{Wang13,Frisenda18, Fan20}, and will contribute to accelerate research in 2DMs. It permits the dry pick-up of 2D crystals directly from SiO$_2$ substrates and to transfer them with precise $x-y-\theta$ alignment. A complete transfer takes under 60 minutes and has a success rate around 95$\%$. Most distinctively, this method can be used to transfer suspended ultra-thin materials such as monolayer and bilayer graphene over areas $\geq$ 10 $\mu$m$^2$, and suspension heights as low as 160 nm without critical point drying. Careful Raman spectroscopy and optical imaging showed no microscopic disorder, tear or a significant density of bubbles in the transferred materials. We demonstrated the assembly of planar optical cavities able to broadly tune BLG's Raman scattering intensity, by up to 19 folds in on-substrate devices and nearly 4 folds in suspended structures. We calculated that the BLG exclusive light absorption can be engineered by a similar ratio. Our fabrication method fills a major gap in previous transfer methods by making possible a versatile transfer of suspended 2DMs (any material, stacking-order, substrate) all the way down to the ultra-thin limit (SLG, BLG). We foresee that this fabrication route can create heterostructures suited for exploring the interplays of nanoscale mechanics, optics, and electronics, for instance in twistronics \cite{Cao18, Hu2020, Weston20}, straintronics\cite{Naumis17, McRae19, Zhang20}, and optoelectronics \cite{Florian18,Midolo18, Eggleton19, Li19_2, Zhang19_2}.

\acknowledgement
This work was supported by NSERC (Canada), CFI (Canada), and Concordia University. We acknowledge usage of the QNI (Quebec Nano Infrastructure) cleanroom network.

\section{Contributions}

IGR and FCRM made equal contributions. They led the development of the methods, fabrication of the samples, measurements, data analysis, and calculations. ARC conceived and supervised the project, he made contributions to all aspects of the project. WW contributed to sample fabrication. GJM contributed to Raman measurements and data analysis. IGR, FCRM, and ARC wrote the manuscript, and all authors reviewed the manuscript.

\section{Associated Content}
Supporting Information: Detailed step-by-step description of the 2DM transfer method, information about all on-substrate heterostructures transferred, information about all suspended heterostructures transferred, description of the theoretical calculations of the Raman factors (and light absorption) and their comparison to the experimental data. (PDF)
\\
Supporting Movie: Video of the step-by-step 2DM transfer procedure.

\bibliography{Rebollo_Rodrigues_stamping_revision1}

\end{document}


\BeginNoToc
\begin{center}
\section*{\Large{Supporting Information}}
\end{center}
\EndNoToc

\tableofcontents

\section{Deterministic any-stacking-order and suspended 2DM transfer procedure}\label{sec:S1}

The stamping method is separated into five steps as shown in Figure 1a-e of the main text. The first step consists in the preparation of a nitrocellulose micro-stamp and finding a desired 2DM crystal on a \ch{SiO2} substrate. Secondly, the micro-stamp is aligned above the 2DM and brought into contact. Later, the 2DM is picked-up from the \ch{SiO2} substrate by carefully retracting the micro-stamp. The micro-stamp/2DM assembly is subsequently aligned above the new substrate and brought into contact. The transfer is completed with a facile microliter-volume solvent dissolution of the micro-stamp. We summarize the key details of each of these five stages below, and we first briefly describe our stamping apparatus.

\subsection{Stamping apparatus with $x, y, z, \theta$ alignment}\label{sec:S1.1}

\begin{figure}
	\includegraphics[scale=1]{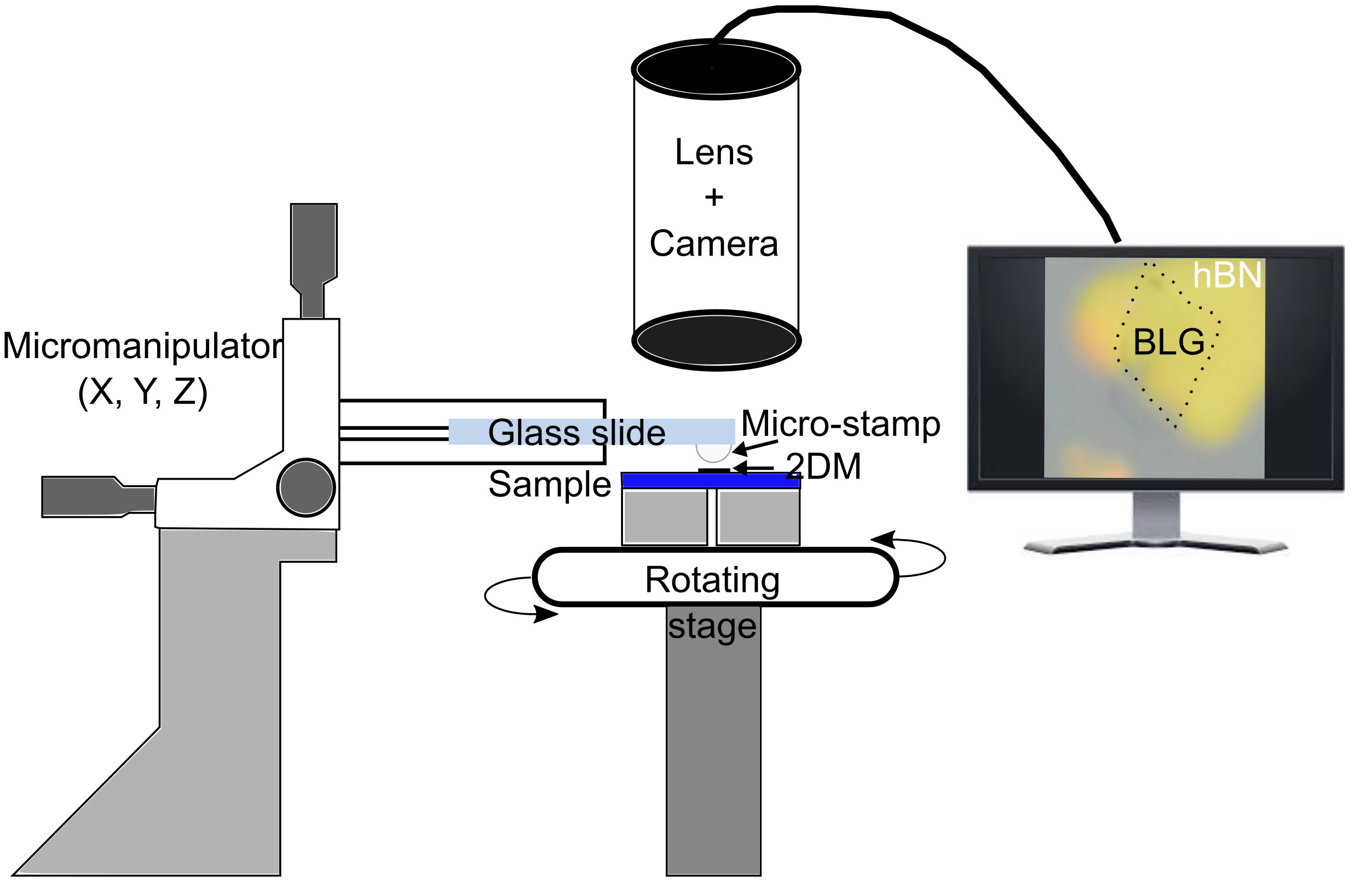}
	\caption[Stamping apparatus]{Stamping apparatus. The substrate with the 2DM-to-be-transferred is placed on a rotating stage and held by vacuum, while a glass slide with a custom micro-stamp is attached to the x-y-z manipulator. A long working distance optical assembly is connected to a digital camera which displays a live view of the transfer process.}
	\label{fig:S1}
	\end{figure}

The stamping setup used is shown in \Cref{fig:S1}. It consists of a rotating stage with a vacuum system to hold a substrate in place, a three-axis micro manipulator ($x, y$ and $z$-axis) that holds a glass slide with the micro-stamp (nitrocellulose droplet), a long working distance optical assembly, and a high-resolution CCD camera which is connected to a monitor for live viewing. This setup is based on previous stamping techniques, often called deterministic transfer \cite{Castellanos14,Frisenda18,Tao18,Hemnani18}. Most steps of our micro-stamp transfer process are all-dry, and only the last step requires micro-liter amounts of mild solvents.

\subsection{Micro-stamp preparation and 2DM exfoliation}\label{sec:S1.2}

\begin{figure}
	\includegraphics[scale=1]{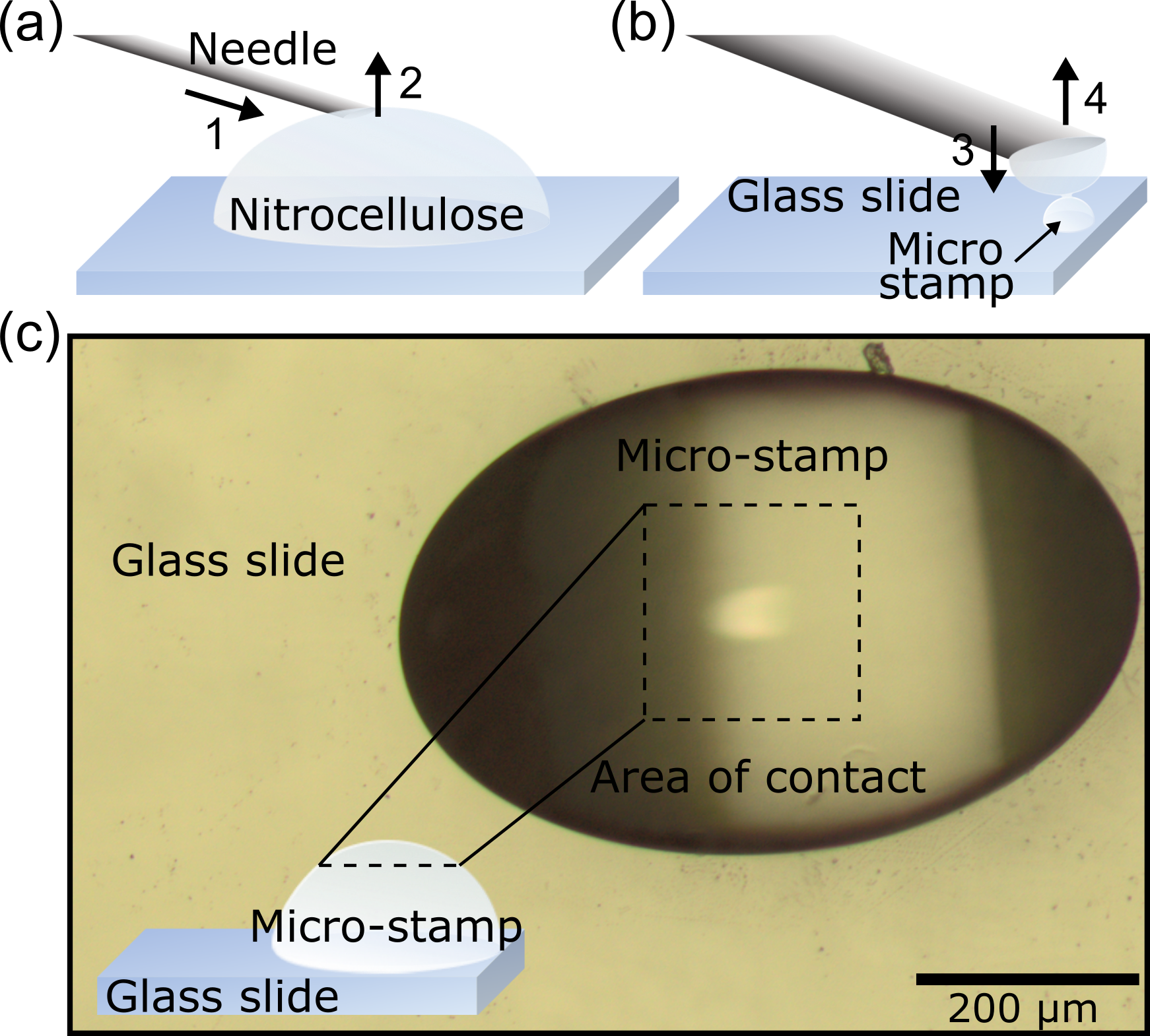}
	\caption[Schematic of the nitrocellulose micro-stamp preparation]{Nitrocellulose micro-stamp preparation. (a) A sharp needle is submerged in a drop of nitrocellulose-solution (Extra Life$^{\mathrm{TM}}$ No Chip Top Coat -- Revlon), and then pulled up so that a small droplet remains attached to the needed by capillary forces. (b) A slight contact of the droplet with the target glass slide leaves a reproducible micro-stamp. (c) Top-view optical image of a typical micro-stamp, with a 200 $\mu$m $\times$ 200 $\mu$m contact area for 2DM transfer. Inset: Illustration of the micro-stamp profile.}
	\label{fig:S2}
	\end{figure}

\textit{Micro-stamp preparation} -- Previous deterministic 2DM transfer methods have used stamps based on materials such as polydimethylsiloxane (PDMS) \cite{Castellanos14,Tao18}, polymethylmethacrylate (PMMA) \cite{Uwanno15}, thermal release tape (TRP) \cite{Kim15}, polyvinyl alcohol (PVA) \cite{Frisenda18}, and polypropylene carbonate (PPC) \cite{Kim16,Jung19}. The novelty of our stamping method lies in the use of a nitrocellulose-based stamp in the shape of an ellipsoidal micro-droplet (see \Cref{fig:S2}). The stamp is made with a commercially available product (Extra Life$^{\mathrm{TM}}$ No Chip Top Coat -- Revlon). The size and shape of the micro-stamp determine the contact area between the stamp and 2DMs. The optimal stamp size for our work was around 400 $\mu$m $\times$ 600 $\mu$m $\times$ 400 $\mu$m (see \Cref{fig:S2}), with a contact area (top of stamp) approximately 200 $\mu$m by 200 $\mu$m. To achieve these micro-stamp dimensions, we first submerge the tip of a 27 gauge needle into a small drop of the solution deposited on a glass slide (\Cref{fig:S2}a). Due to capillary forces, a small volume of solution stays on needle when pulled away from it. When the needle barely touches the target clean glass slide, it transfers a very smal droplet on its surface (see \Cref{fig:S2}b). The resulting micro-stamp is inspected by optical microscopy to make sure that it has the wanted shape and size. The narrow apex of the stamp appears as a bright spot in \Cref{fig:S2}c. This process only takes a few minutes and is easily reproducible.

\noindent
\textit{2DM exfoliation} -- To exfoliate graphene crystals, we start with high quality HOPG graphite flakes (Kish Graphite B from Covalent Materials Corporation). To exfoliate hexagonal boron-nitrate (hBN) 2D crystals, we start with high quality hBN crystals (PT110 Powder CTR from Momentive Performance Materials). To exfoliate molybdenum disulfide (\ch{MoS2}) crystals, we start with a synthetic \ch{MoS2} crystal from 2D Semiconductors. Our wafers are made of 500 $\mu$m-thick Si with a 310 nm-thick \ch{SiO2} film grown on them. We transferred a photolithography patterned coordinate grid on the SiO$_2$/Si wafers, to easily record the location of the candidate 2DM crystals. To exfoliate 2DMs, we first do a coarse mechanical cleavage with a razor blade to generate the thinnest flakes possible. We then place these flakes on a piece of Scotch$^{\mathrm{TM}}$ tape (exfoliation of hBN is done using Nitto$^{\mathrm{TM}}$ tape) and further cleavage is done by repeatedly folding the tape \cite{Bonaccorso12}.

Before transferring the exfoliated 2DM from the tape to the \ch{SiO2}/Si substrate, we lightly etch the substrate with a dilute hydrochloric acid (HCl) and hydrogen peroxide (H$_2$O$_2$) solution (18:1:1) at 75 C$^0$ for 5 minutes, to get rid of any surface residues. This etching minimizes the substrate contamination and promotes adhesion between 2DM crystals and the oxide. The substrate is carefully rinsed with DI water, blown dry with nitrogen and baked at 120 $^0$C for 2 minutes to evaporate any water left. The tape containing the 2DM flakes is gently pressed with a finger on the substrate, and we wait for about 10 minutes before slowly peeling off the tape (0.1 mm/s) with the help of tweezers. The scotch tape itself leaves organic residues that need to be cleaned before the deterministic transfer. The substrate with exfoliated 2DM is submerged in a warm bath of acetone at 75 $^0$C for 5 minutes, rinsed with IPA, then with DI water and baked at 120 $^0$C for 2 minutes. The substrate is then mounted on the vacuum stage of the stamping apparatus (\Cref{fig:S1}) and the lens and camera system is focused on the target crystal.

\subsection{Making contact between the micro-stamp and 2D crystal}\label{sec:S1.3}

The glass slide with the micro-stamp is mounted on the micro-manipulator, the apex of the stamp is centred 1 mm above the crystal selected for pick-up (the micro-stamp is highly transparent and acts as a lens due to its ellipsoidal shape). The stamp is left to dry for 10 $\pm$ 3 min. During this time the stamp's surface hardens. It is then lowered slowly ($\sim$ 50 $\mu$m/s) to contact the selected crystal and its immediate surrounding area (4000 $\mu$m$^2$). There is a sudden change of colour when contact is made (see Figure 1b in the main text). The contact must be done on the first trial, otherwise the micro-stamp deforms, and the pick-up process may induce crystal folding or work unreliably. The stamp-2DM contact is maintained constant for 20 $\pm$ 5 min to promote strong adhesion. These parameters were the same for all of the 2D crystals we transferred: SLG, BLG, FLG, hBN and \ch{MoS2} crystals.

\subsection{Direct pick-up from \ch{SiO2}}\label{sec:S1.4}

\begin{figure}
	\includegraphics[scale=1]{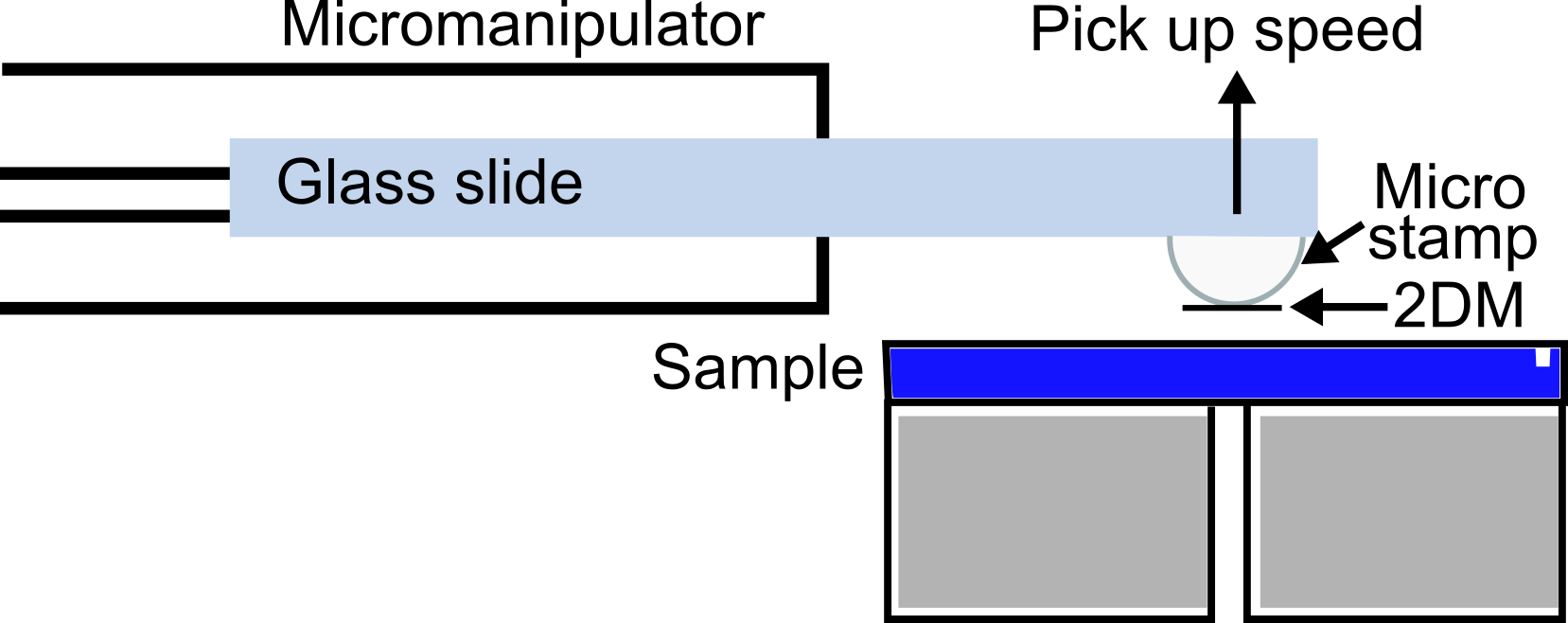}
	\caption[Schematic of 2DM pick-up from \ch{SiO2} substrate]{Schematic of the 2DM pick-up process from a \ch{SiO2} substrate. A micro-manipulator holds the glass slide to which the micro-stamp is attached. By raising the slide, the 2DM is cleanly picked-up.}
	\label{fig:S3}
	\end{figure}

The pick-up speed of the micro-stamp/2DM assembly away from the original substrate is controlled with the $z$-axis of the micro-manipulator. The optimal speed depends on the thickness of the crystal. We found that the nitrocellulose-based stamp acts as a hard surface at higher speeds and as a more flexible one at lower speeds. The vertical pick-up speed used for atomically thin crystals (SLG, BLG and few layers) is $\sim$ 500 $\mu$m/s, while for thicker crystals it is reduced to $\sim$ 250 $\mu$m/s. The live view option of the camera allows us to observe when the crystal is completely picked-up and if the process induces crystal folds. Once the crystal is picked up (see \Cref{fig:S3}), we raise the stamp by an additional 500 $\mu$m, and exchange the old \ch{SiO2}/Si substrate with the new target substrate.

\subsection{Alignment and stamping of a 2D crystal on a target substrate}\label{sec:S1.5}

\begin{figure}
	\includegraphics[scale=1]{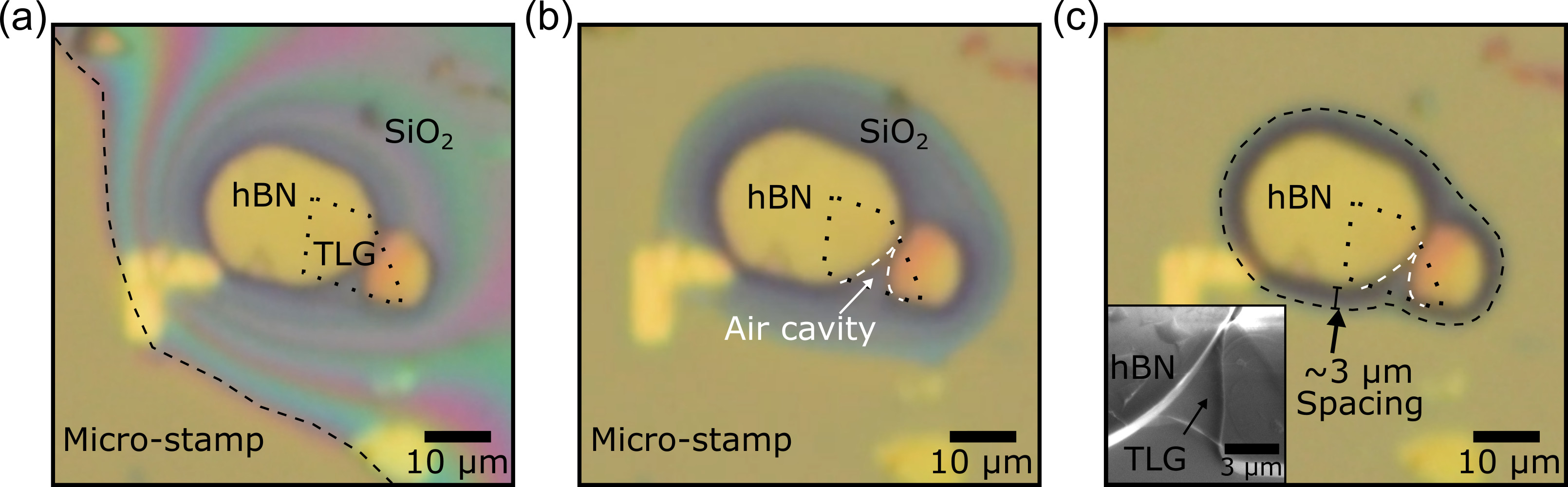}
	\caption[Alignment and stamping of 2DM]{Alignment and stamping of a trilayer (TLG) graphene crystal on a hBN trench. (a) The micro-stamp starts to make contact with the target substrate. (b) TLG crystal makes slight contact with hBN substrate. (c) The micro-stamp holding the TLG crystal is in contact with the \ch{SiO2} and hBN, but does not contact a 3 $\mu$m-wide zone surrounding the hBN support. Inset: Tilted SEM image of the final heterostructure.}
	\label{fig:S4}
	\end{figure}

We often use hBN as the target substrate for transfer, but also successfully used \ch{SiO2}, graphene, or aluminum. We note that exfoliated hBN crystals on \ch{SiO2}/Si substrates often exhibit ``U'' shape edges which can serve as naturally occurring trenches to suspend 2DMs. We first characterize the substrate via AFM (thickness, shape, flatness). Then, we place it on the stamping stage, and focus the optical assembly on the substrate. We mark (trace) the substrate crystal (e.g. hBN) contour on the video screen for future reference. The 2DM-substrate alignment is done in successive steps. First, the micro-stamp is centred and positioned about 0.5 mm above the target, and the substrate is aligned as desired relative to the 2D crystal orientation on the micro-stamp. In a second stage, we focus the image right above the substrate and lower progressively the 2D crystal at about 20 $\mu$m/s until both the substrate and 2D crystal are clearly in focus. In \Cref{fig:S4} we show a final alignment done while making gentle contact at about 5 $\mu$m/s. We stop lowering the micro-stamp when the 2D crystal contacts the target, as shown in \Cref{fig:S4}b. We ensure that the micro-stamp is not pressed hard enough to deform significantly near its apex (stamping area), and does not contact the SiO$_2$ immediately surrounding the hBN substrate. As visible in \Cref{fig:S4}c, there should be a spacing (a no-contact zone) of $\sim$ 3 $\mu$m between the stamp (in contact with \ch{SiO2}) and the hBN substrate. A tilted SEM image of a suspended TLG structure can be seen in the inset of \Cref{fig:S4}c.

\subsection{Transfer: Solvent dissolution of micro-stamp}\label{sec:S1.6}

\begin{figure}
	\includegraphics[scale=1]{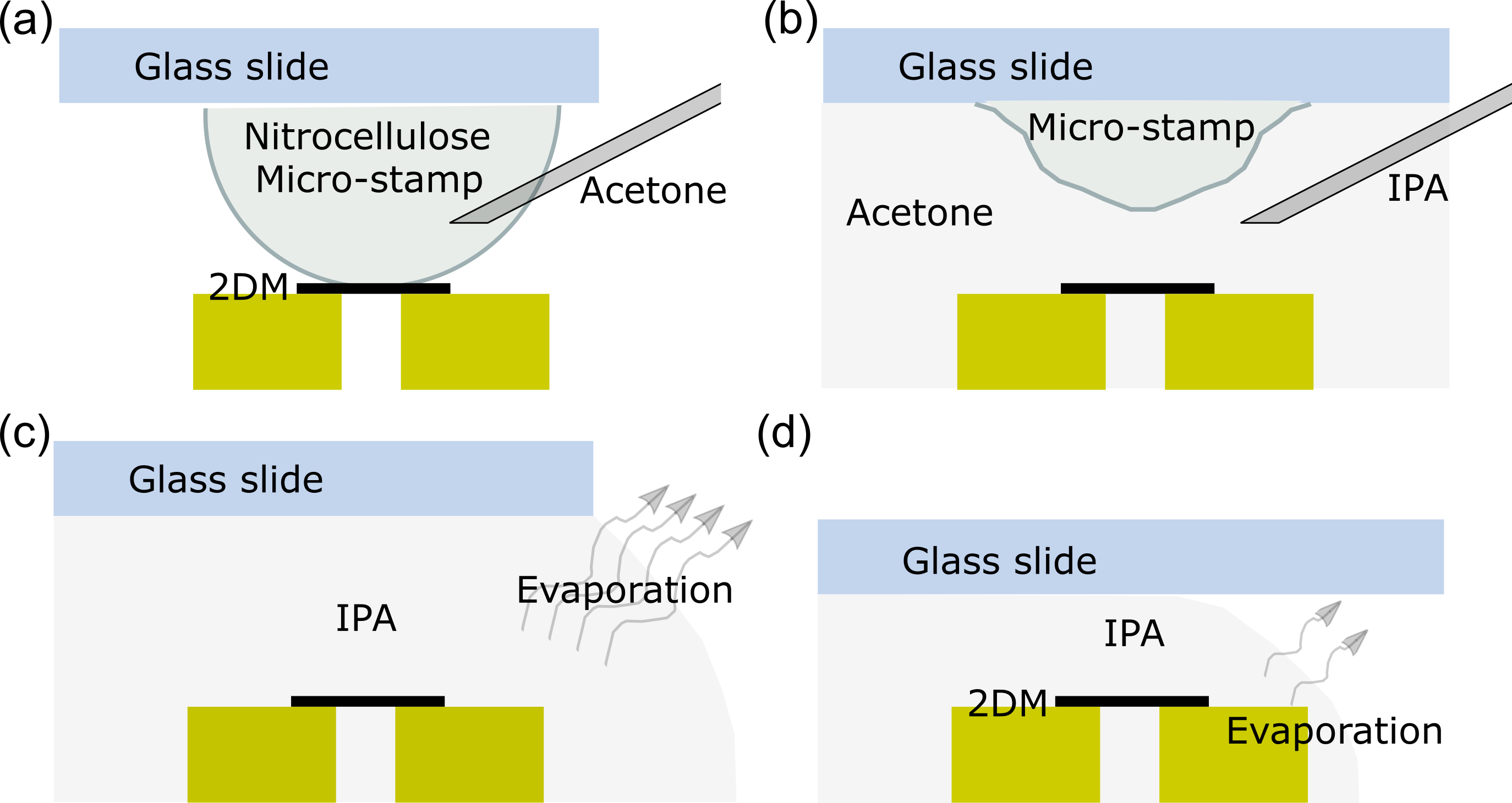}
	\caption[2DM Transfer via solvent dissolution of micro-stamp]{2DM transfer via solvent dissolution of the micro-stamp (a) A micro-pipette is used to introduce a drop of acetone. (b) The acetone starts to dissolve the stamp and the 2D crystal is released. A few IPA drops are injected to flush the acetone and polymer residues. (c) The stamp is completely dissolved, and fast evaporation of the solvent starts to occur. (d) By the lowering of the glass slide we can decrease the rate of evaporation, minimizing the risk of collapse for suspended crystals.}
	\label{fig:S5}
	\end{figure}

A clean micro-pipette is used to inject one drop of acetone ($<$ 500 $\mu$L) in the spacing between the micro-stamp and the substrate (\Cref{fig:S5}a). The acetone rapidly dissolves the stamp (this can be monitored in real time on the screen) and releases the 2D crystal. Once the stamp has been dissolved, we raise the glass slide by 500 $\mu$m and do a local rinsing with IPA using the same micro-pipette. We repeat a few times this rinsing with IPA to completely flush the acetone and polymer residues (\Cref{fig:S5}b). At this point the 2D crystal has been transferred on the new substrate and is surrounded by the IPA solution (\Cref{fig:S5}c). To transfer suspended crystals, we control the evaporation rate of IPA to avoid their collapse due to capillary forces. The evaporation rate is easily tuned by raising or lowering the glass slide to tailor the exposure of the IPA solution to the atmosphere (\Cref{fig:S5}d). This procedure removes the need for critical point drying of our suspended 2DMs. It enables the stamping of defect-free suspended crystals with a simple table-top apparatus.

\section{On-substrate heterostructures}\label{sec:S2}

In Figure 2 of the main text, we describe the capabilities of our method to transfer naked (not encapsulated) 2DM crystals. We fabricated on-substrate heterostructures with precise alignment of large 2D crystals. Below, we present in Table \ref{tab:S1} a detailed list of our 21 attempts at transferring 2DM crystals from one substrate to another following the recipe presented in Section \ref{sec:S1} and Figure 1 of the main text.

	\begin{figure}
	\includegraphics[scale=1]{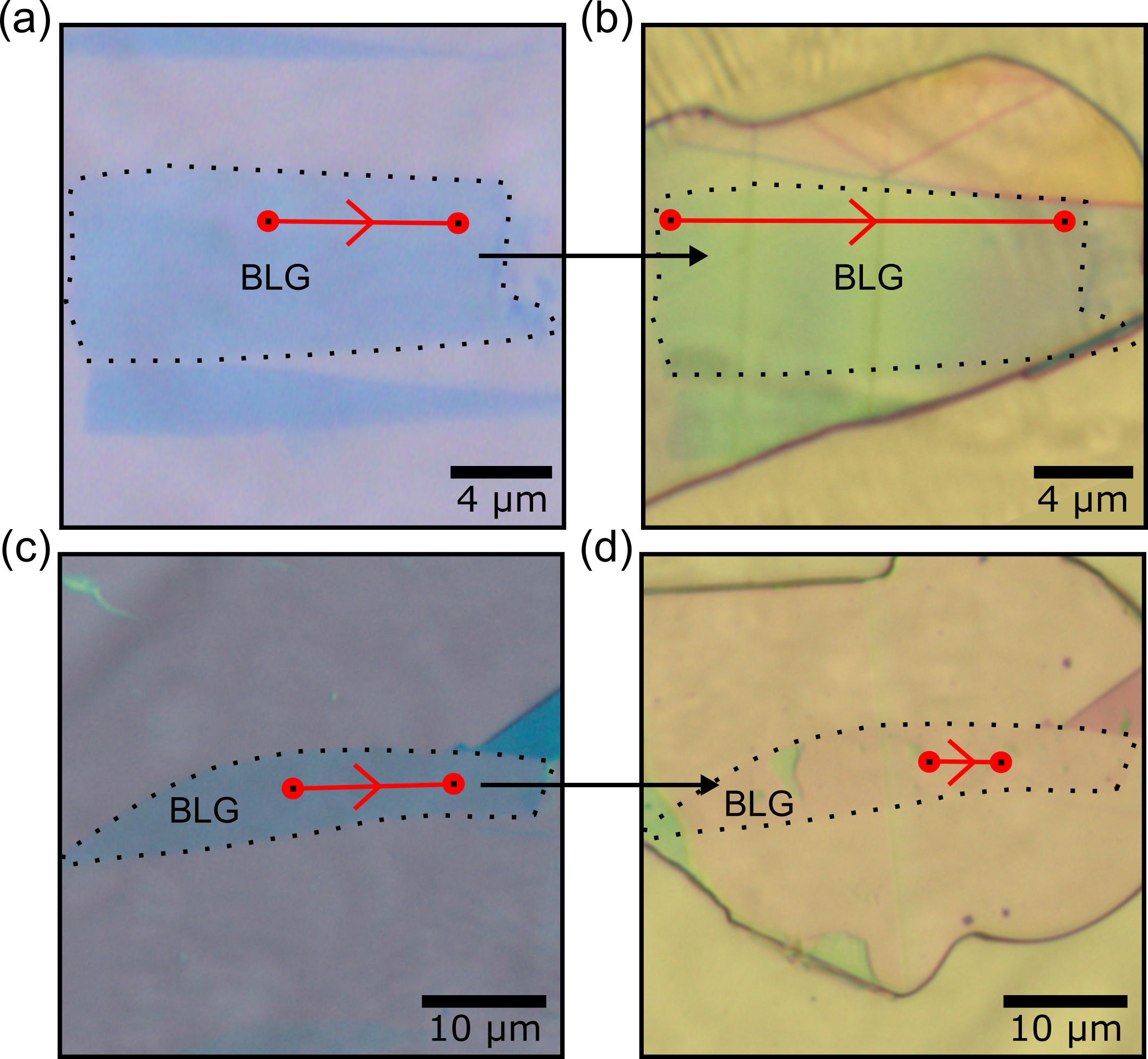}
	\caption[Transferring large bilayer graphene (BLG) crystals from \ch{SiO2} substrate to hBN substrates]{Transferring large bilayer graphene (BLG) crystals from \ch{SiO2} substrates to hBN substrates. (a) Optical image of a large BLG crystal on \ch{SiO2} before the transfer procedure, and (b) the same BLG crystal on hBN after the transfer. The BLG was picked up by the micro-stamp and transferred, without any tearing or encapsulation of the BLG. The red lines show the location of Raman spectra acquisitions. (c)-(d) Same information as in (a)-(b) for another large area BLG crystal transfer.}
	\label{fig:S6}
	\end{figure}
		
For each device listed in Table \ref{tab:S1}, we acquired high quality optical images of the 2DM before and after its transfer, as shown in \Cref{fig:S6} for two large area BLG crystals. These optical images were used to extract the surface area of the samples listed in Table \ref{tab:S1}, and to verify that no significant macroscopic damage (tearing, folding, and bubbles) was introduced during the transfer. Before and after transfer, we acquired Raman scans (along the red lines in \Cref{fig:S6}) and AFM images of the devices. The comparison of the Raman spectra (Figure 2) before and after transfer confirmed that no microscopic disorder was introduced during the transfer. The alignment accuracy of the transfer is $\approx$ 1 $\mu$m when great care is taken. We successfully transferred all of the 2DMs we attempted to. They include various thicknesses of graphene (MLG, BLG, TLG, FLG), \ch{MoS2}, and hBN. \Cref{tab:S1} details the thickness of the materials transferred and the final heterostructure geometries created.

	\strutlongstacks{T}
	\begin{table} \setlength{\tabcolsep}{6pt}
	\begin{tabular}{ c c c c c }
 	\toprule[1pt]\midrule[0.3pt]
 	    \multirow{3}*{ \Centerstack{Device \\name} }
	& \multicolumn{2}{c}{2D crystal area ($\mu$m$^2$)}
	& \multirow{3}*{ \Centerstack{Transferred  \\ material and \\ thickness} }
	& \multirow{3}*{ \Centerstack{Final heterostructure} }
	\\ \cmidrule(lr){2-3}
	& \multirow{2}*{Pre-transfer} 	& \multirow{2}*{Post-transfer}	& & \\
	&&&&
	\\ \midrule
 	SA1	&	15	&	15	&	SLG	&	SLG/hBN/Al	\\ \midrule[0.2pt]
 	SA2	&	25	&	25	&	SLG	&	SLG/hBN/\ch{SiO2}/Si	\\ \midrule[0.2pt]
	BA1	&	140	&	140	&	BLG	&	BLG/hBN/Al	\\ \midrule[0.2pt]
	BA2	&	50	&	25	&	BLG	&	BLG/hBN/Al	\\ \midrule[0.2pt]
	BA3	&	60	&	15	&	BLG	&	BLG/hBN/Al	\\ \midrule[0.2pt]
	BA4	&	190	&	190	&	BLG	&	BLG/hBN/Al	\\ \midrule[0.2pt]
	BA5	&	20	&	120	&	BLG	&	BLG/hBN/Al	\\ \midrule[0.2pt]
	BA6	&	40	&	40	&	BLG	&	BLG/hBN/Al	\\ \midrule[0.2pt]
	BAX	&	90	&	0	&	BLG	&	failed		\\ \midrule[0.2pt]
	TA1	&	350	&	350	&	TLG	&	TLG/hBN/\ch{SiO2}/Si	\\ \midrule[0.2pt]
	FA1	&	50	&	50	&	FLG	&	FLG/hBN/Al	\\ \midrule[0.2pt]
	FA2	&	360	&	360	&	FLG	&	FLG/hBN/Al	\\ \midrule[0.2pt]
	FA3	&	670	&	670	&	FLG	&	FLG/hBN/\ch{SiO2}/Si	\\ \midrule[0.2pt]
	FA4 &   410 &   410 &   FLG &   FLG/\ch{SiO2}/Si	\\ \midrule[0.2pt]
    MoA1&	290	&	290	&\ch{MoS2}  -- few layers&\ch{MoS2}/hBN/\ch{SiO2}/Si	\\ \midrule[0.2pt]
	MoA2&	120	&	120	&\ch{MoS2}  -- 10nm&	\ch{MoS2}/hBN/\ch{SiO2}/Si	\\ \midrule[0.2pt]
	BNA1&	1150	&	1150	&	hBN -- 110nm	&	hBN/\ch{SiO2}/Si	\\ \midrule[0.2pt]
	BNA2&	580	&	580	&	hBN -- 610nm	&	hBN/\ch{SiO2}/Si	\\ \midrule[0.2pt]
	BNA3&	450	&	450	&	hBN -- 960nm	&	hBN/\ch{SiO2}/Si	\\ \midrule[0.2pt]
	BNA4&	550	&	550	&	hBN -- 1250nm	&	hBN/\ch{SiO2}/Si	\\ \midrule[0.2pt]
	BNA5&	290	&	290	&	hBN -- 1500nm	&	hBN/\ch{SiO2}/Si	\\
 	\midrule[0.3pt]\bottomrule[1pt]
	\end{tabular}
    \captionsetup{justification=centering, type=table,position=top}
	\caption{\label{tab:S1}List of all 21 on-substrate 2DM transfers done via our nitrocellulose micro-stamp method.}
	\end{table}

The transfer procedure was very reproducible, and its success rate was very high. Out of the 21 attempts, there was only 1 failure (no transfer) and 2 partial successes (some tearing of the crystal), and 18 fully successful transfers.  In terms of producing the desired planar heterostructures, we thus find the success rate to be around 95\%.

\section{Suspended heterostructures}\label{sec:S3}

In Figure 3 of the main text, we describe our capability to transfer and align suspended ultra-thin 2DM crystals on hBN trenches. We created suspended SLG, BLG, and FLG heterostructures with various suspension heights, including some structures with a variable (tilted) suspension height as shown in Figure 4. Below, we present in \Cref{tab:S2} a detailed list of our 15 attempts at transferring 2DM crystals from \ch{SiO2} substrates onto hBN trenches for suspension. We followed the recipe presented in \Cref{sec:S1} and Figure 1 of the main text.

\begin{figure}
	\includegraphics[scale=1]{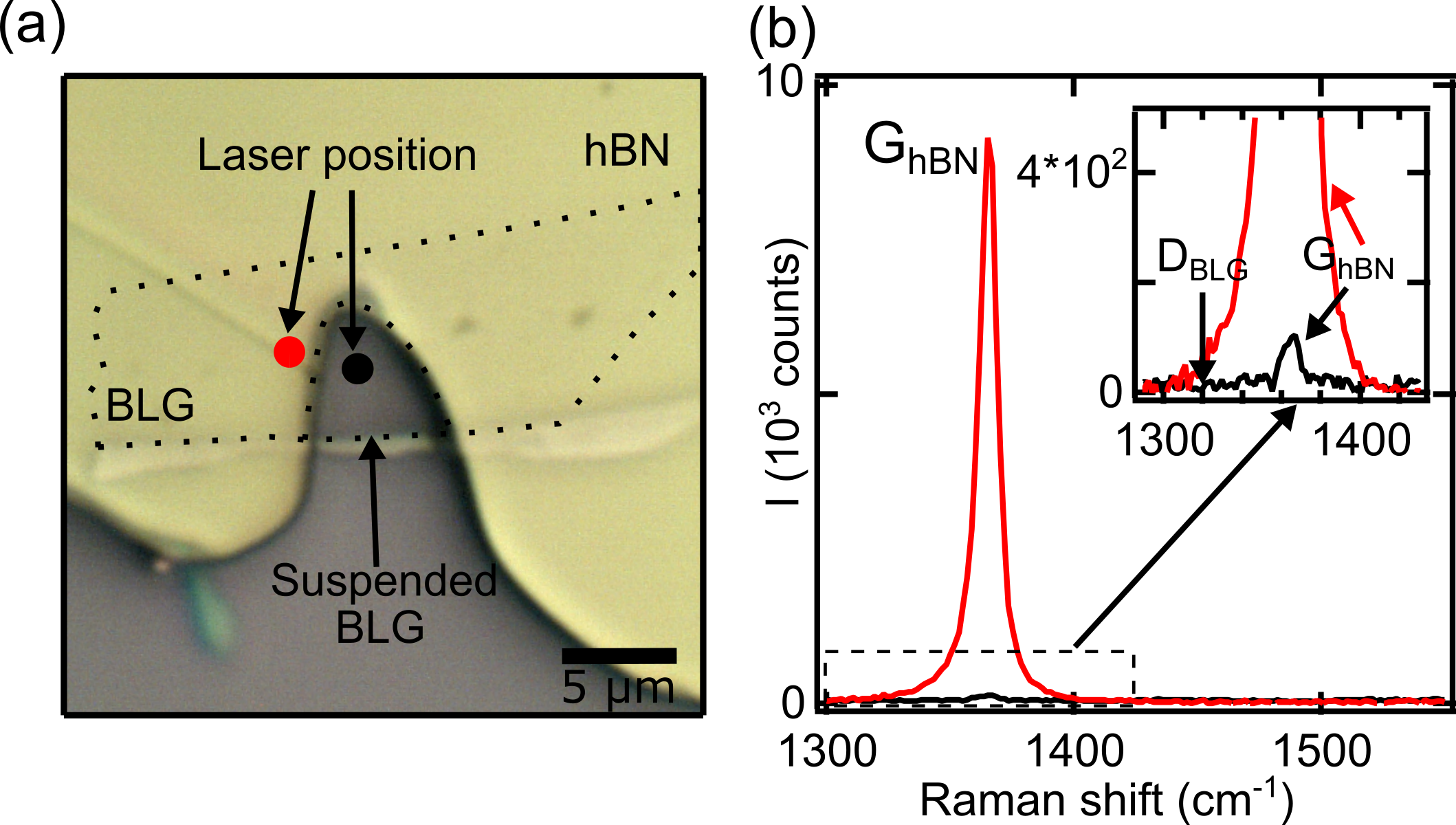}
	\caption[Transferred suspended bilayer graphene heterostructures]{Comparing the \GhBN peak intensities at different laser positions in a suspended BLG heterostructure. (a) Top view image with a red-labeled and a black-labeled laser positions, respectively located on the hBN-supported and suspended BLG regions. (b) BLG Raman data acquired at the location where the crystal is suspended (black data) and where it is supported by a hBN substrate (red data). The inset shows a zoom-in on the data.}
	\label{fig:S7}
	\end{figure}

Pristine graphene crystals feature two main Raman resonances, a first order G-peak and higher order 2D-peak. In defected graphene crystals, there is a third Raman resonance called D-peak which appears. All of these resonances are laser wavelength dependent. With a 532-nm laser wavelength, the location of the D peak is around 1320 cm$^{-1}$ \cite{Wu18}, G around 1580 cm$^{-1}$ \cite{Wu18}, and 2D around 2700 cm$^{-1}$. Our heterostructures often include hBN crystals beneath the graphene crystals, and this material is also Raman active. It features a resonance called \GhBN located around 1370 cm$^{-1}$ \cite{Lee19}. It is possible to clearly distinguish between a graphene D-peak, and a \GhBN. In Figure 3 of the main text, we show suspended graphene crystals on top of ``U'' shaped hBN trenches. There is a small portion of incoming laser photons being scattered by the edge of the hBN trench, and this leads to a very small intensity \GhBN peak. In \Cref{fig:S7}, we compare the intensities of the \GhBN peak at different laser positions in the device shown in Figure 3c.

\begin{figure}
	\includegraphics[scale=1]{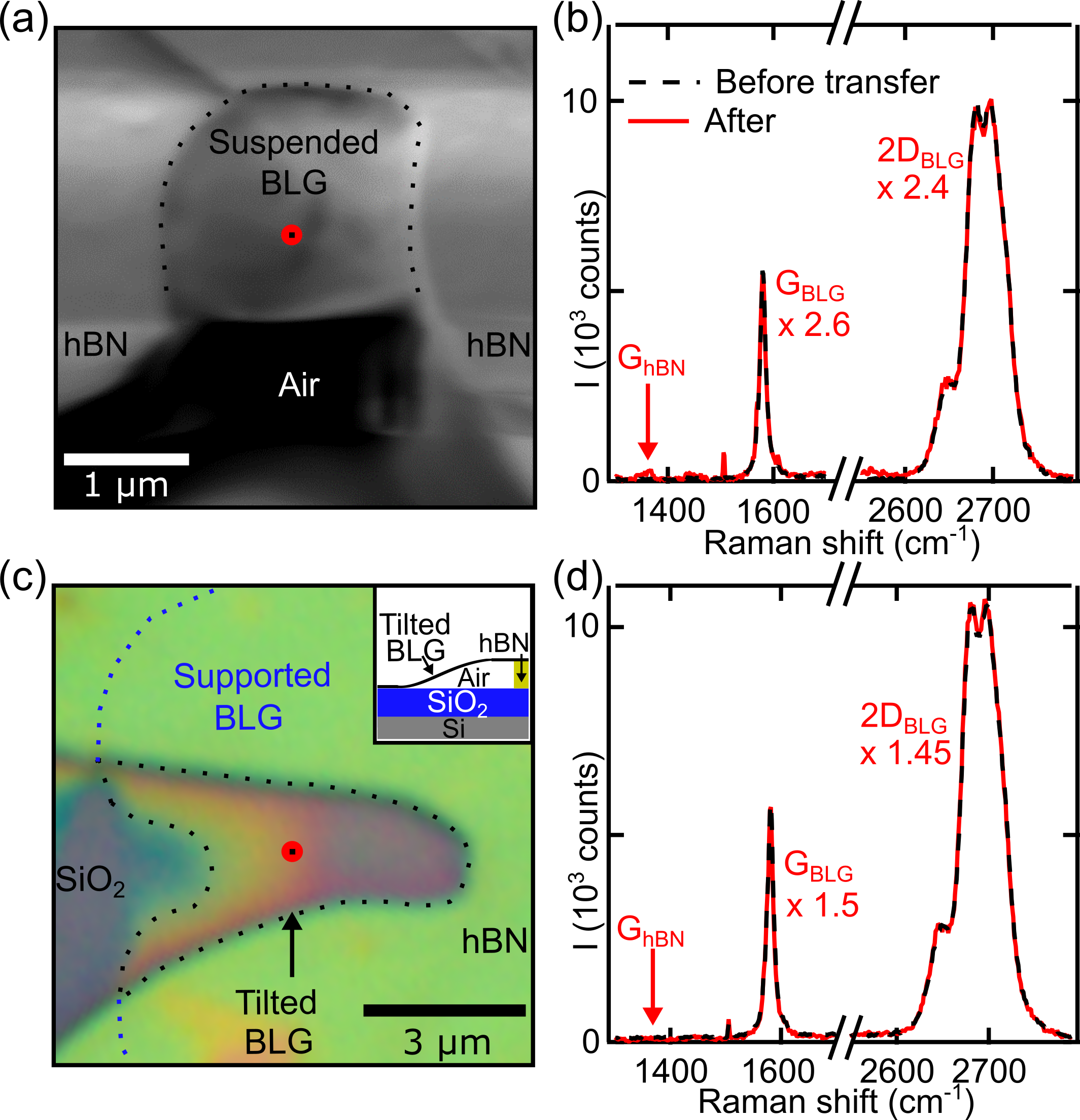}
	\caption[\textcolor{red}{Transferred suspended BLG}]{Transferred suspended BLG. (a) SEM image of a transferred suspended BLG crystal with a uniform height. (b) Raman spectra acquired at the red-marker location in (a), before (black data) and after (red data) the BLG transfer. (c) Optical image of a transferred tilted-suspended BLG crystal. (d) Raman spectra acquired at the red-marker location in (c), before (black data) and after (red data) the BLG transfer.}
	\label{fig:S8}
	\end{figure}

None of our fabricated samples featured a defect-related D peak. In \Cref{fig:S8}, we show two additional transferred suspended BLG devices and their Raman spectra. A Raman D-peak is not visible, and neither is the \GhBN peak. This complete absence of \GhBN is because the laser beam was farther away from the hBN edge during the Raman acquisition than for the device in \Cref{fig:S7}.

For each device listed in \Cref{tab:S2}, we acquired high quality optical images of the 2DM before and after the transfer. We also acquired SEM images after suspension in some devices, as shown in \Cref{fig:S8} for two suspended BLG heterostructures. One of these two devices has a uniform suspension height (\Cref{fig:S8}a) while the other has a titled-suspension (\Cref{fig:S8}b) resulting in a continuous variation of its suspension height. These images, and others, were used to extract the surface area of the crystals listed in \Cref{tab:S2}, and to verify that no significant macroscopic damage (tearing, folding, and bubbles) was introduced during the transfer. Before and after the transfer we also acquired multiple Raman scans of the samples. The Raman spectra, see \Cref{fig:S8} and Figure 3, confirmed that no microscopic disorder was introduced during the stamping. The translational and rotational alignment accuracy of the deterministic transfer were down to 1 $\mu$m and less than 1 degree. We successfully transferred the materials we attempted to, including various thicknesses of graphene (SLG, BLG, TLG, FLG). The suspended heterostructures created were graphene/air/\ch{SiO2}/Si with hBN trenches.

\begin{figure}
	\includegraphics[scale=1]{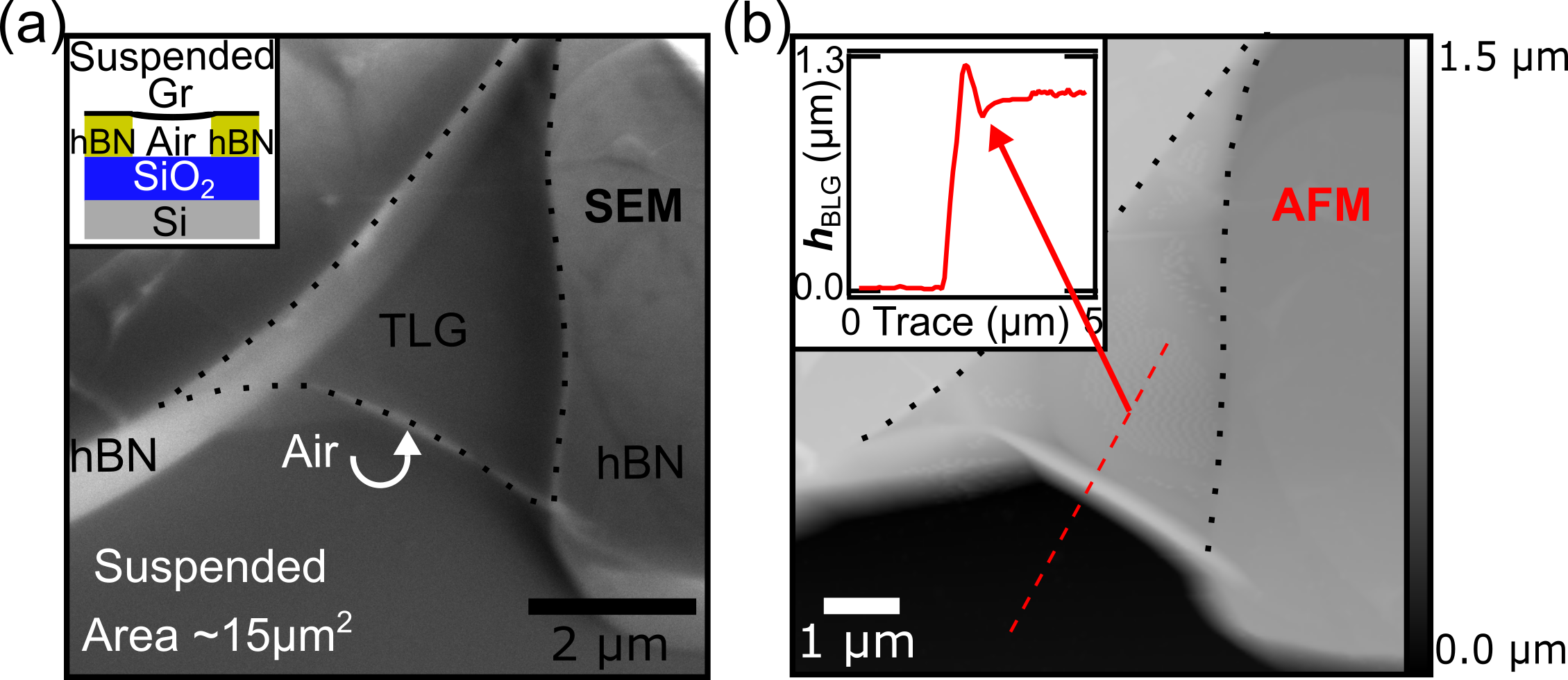}
	\caption[Transferred suspended TLG heterostructure]{Transferred suspended TLG heterostructure. (a) Tilted-SEM image showing a uniform and wrinkle-free suspension. Inset: diagram of the heterostructure geometry. (b) AFM image of the same suspended TLG device. The inset shows the AFM data along the dashed-red line cut in the main panel. }
	\label{fig:S9}
	\end{figure}

In order to determine the suspension height of graphene, and verify the absence of wrinkles introduced during transfer, we used tilted-SEM imaging and AFM imaging, as shown in \Cref{fig:S9}. The results show uniform suspension heights (except in the tilted devices), and these precise suspension heights were confirmed by the quantitative measurements and modeling of Raman interferences presented in Figure 4. Indeed, the Raman factors are highly sensitive to the suspension height of the graphene \cite{Wu18}, as discussed in the next section.

The suspension transfer procedure is very reproducible. Out of our 16 transfer attempts, there was only 1 failure (transferred, but no suspension) and 15 successes. In terms of producing the desired suspended heterostructures, we thus find the success rate to be around 93\%.

	\strutlongstacks{T}
	\begin{table} \setlength{\tabcolsep}{3pt}
	\begin{tabular}{ c c c c c c c }
 	\toprule[1pt]\midrule[0.3pt]
 	    \multirow{3}*{ \Centerstack{Device \\name} }
	& \multicolumn{3}{c}{2D crystal area ($\mu$m$^2$)}
	& \multirow{3}*{ \Centerstack{Transferred  \\ material} }
	& \multirow{3}*{ \Centerstack{Final \\ heterostructure} }
	& \multirow{3}*{ \Centerstack{Suspension \\ height \\(nm)} }
	\\ \cmidrule(lr){2-4}
	& \small{Pre-transfer} 	&\multicolumn{2}{c}{\small{Post-transfer}} 	& & &
	\\ \cmidrule(lr){3-4}
	& 			&\small{Suspended} & \small{Supported}			& & &
	\\ \midrule
 	SC1	&	90	&	8	&	82	&	SLG	&	SLG/air/\ch{SiO2}/Si	&	550	\\ \midrule[0.2pt]
 	SC2	&	210	&	12	&	198	&	SLG	&	SLG/air/\ch{SiO2}/Si	&	520	\\  \midrule[0.2pt]
 	BC1	&	250	&	12	&	238	&	BLG	&	BLG/air/\ch{SiO2}/Si	&	1150	\\  \midrule[0.2pt]
 	BC2	&	620	&	4	&	616	&	BLG	&	BLG/air/\ch{SiO2}/Si	&	1340	\\ \midrule[0.2pt]
  	BB1	&	600	&	12	&	588	&	BLG	&	BLG/\small air(tilted)\normalsize/\ch{SiO2}/Si&	160	\\ \midrule[0.2pt]	
    BB2	&	370	&	9	&	361	&	BLG	&	BLG/\small air(tilted)\normalsize/\ch{SiO2}/Si	&	360	\\ \midrule[0.2pt]
 	TC1	&	170	&	6	&	164	&	TLG	&	TLG/air/\ch{SiO2}/Si	&	570	\\ \midrule[0.2pt]
 	TC2	&	110	&	15	&	95	&	TLG	&	TLG/air/\ch{SiO2}/Si	&	1050	\\ \midrule[0.2pt]
 	FCX	&	650	&	0	&	650	&	FLG	&	Failed suspension	&	0	\\ \midrule[0.2pt]
 	FC1	&	100	&	8	&	92	&	FLG	&	FLG/air/\ch{SiO2}/Si	&	1300	\\ \midrule[0.2pt]
 	FC2	&	140	&	8	&	132	&	FLG	&	FLG/air/\ch{SiO2}/Si	&	1900	\\ \midrule[0.2pt]
 	FC3	&	160	&	7	&	153	&	FLG	&	FLG/air/\ch{SiO2}/Si	&	1100	\\ \midrule[0.2pt]
 	FC4	&	190	&	3	&	187	&	FLG	&	FLG/air/\ch{SiO2}/Si	&	900	\\ \midrule[0.2pt]
 	FC5	&	100	&	6	&	94	&	FLG	&	FLG/air/\ch{SiO2}/Si	&	1350	\\ \midrule[0.2pt]
 	FC6	&	70	&	7	&	63	&	FLG	&	FLG/air/\ch{SiO2}/Si	&	340	\\ 	\midrule[0.2pt]
    FC7 &   410 &   10  &   400 &   FLG &   FLG/air/\ch{SiO2}/Si	&	1200	\\ 	\midrule[0.2pt]
 	\midrule[0.3pt]\bottomrule[1pt]
	\end{tabular}
	\caption{\label{tab:S2}List of all 16 suspended 2DM transfers done via our nitrocellulose micro-stamp method.}
	\end{table}

\section{Tuning the Raman factors and light absorption in BLG}\label{sec:S4}

\begin{figure}
	\includegraphics[scale=1]{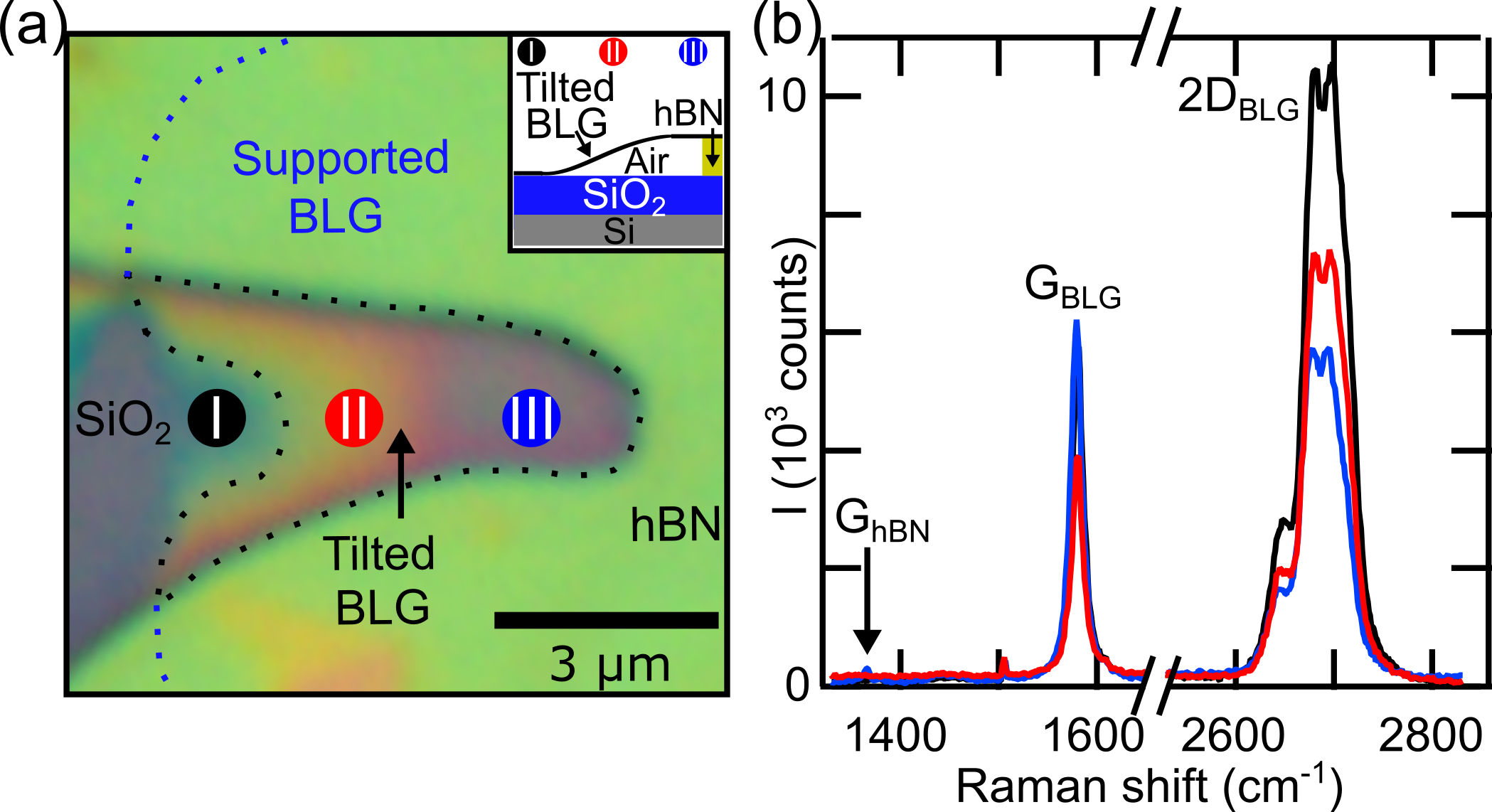}
	\caption[Interference in a BLG heterostructure]{Interference in a tilted-suspended BLG/air/\ch{SiO2}/Si heterostructure. (a) Optical image of tilted-suspended structure with tree distinct regions labeled I, II and III. Inset: diagram of the heterostructure geometry. (b) Raman spectra acquired at the numbered circles location in the optical picture. }
	\label{fig:S10}
	\end{figure}

There is a strong interest in enhancing the light absorption of graphene \cite{Yoon09, Casalino17} and 2D transition metal dichalcogenides (TMDs)\cite{Song15} to optimize their great potential for light harvesting applications, and also to develop new tools for NOEMS research \cite{Midolo18}. The Raman scattering intensity in an isolated (i.e. surrounded by vacuum) 2D crystal is linearly proportional to light absorption since only a tiny fraction of photons undergo Raman scattering \cite{Ferrari13}. In Figure 4 of the main text and \Cref{fig:S10}, we observe that BLG Raman scattering intensity in planar heterostructures can be enhanced via constructive and destructive interferences at the interfaces between the various 2D layers. Here we first present in \Cref{sec:S4.1} a quantitative model based on Fresnel’s equations to calculate the exclusive light absorption in BLG, $A_{BLG}$, and then how this method can be extended to calculate the Raman Factors (Raman relative intensities) $F_{BLG-G}$ and $F_{BLG-2D}$. In \Cref{sec:S4.2} we detail how we analyze and calibrate our experimental Raman data to establish a quantitative comparison between the data and the theoretical model.

\subsection{Calculating the BLG exclusive light absorption and Raman factors in planar heterostructures}\label{sec:S4.1}

\textit{Exclusive light absorption in BLG} -- We use a simple theoretical model based on Fresnel’s equations \cite{Hecht02} and derived in previous work \cite{Casalino17, Song15}. This exclusive light absorption model predicts an absorption in the visible range of about 2.3\% for a SLG (4.6\% for BLG) when surrounded by vacuum. This was confirmed experimentally \cite{Nair08}.

\begin{figure}
	\includegraphics[scale=1]{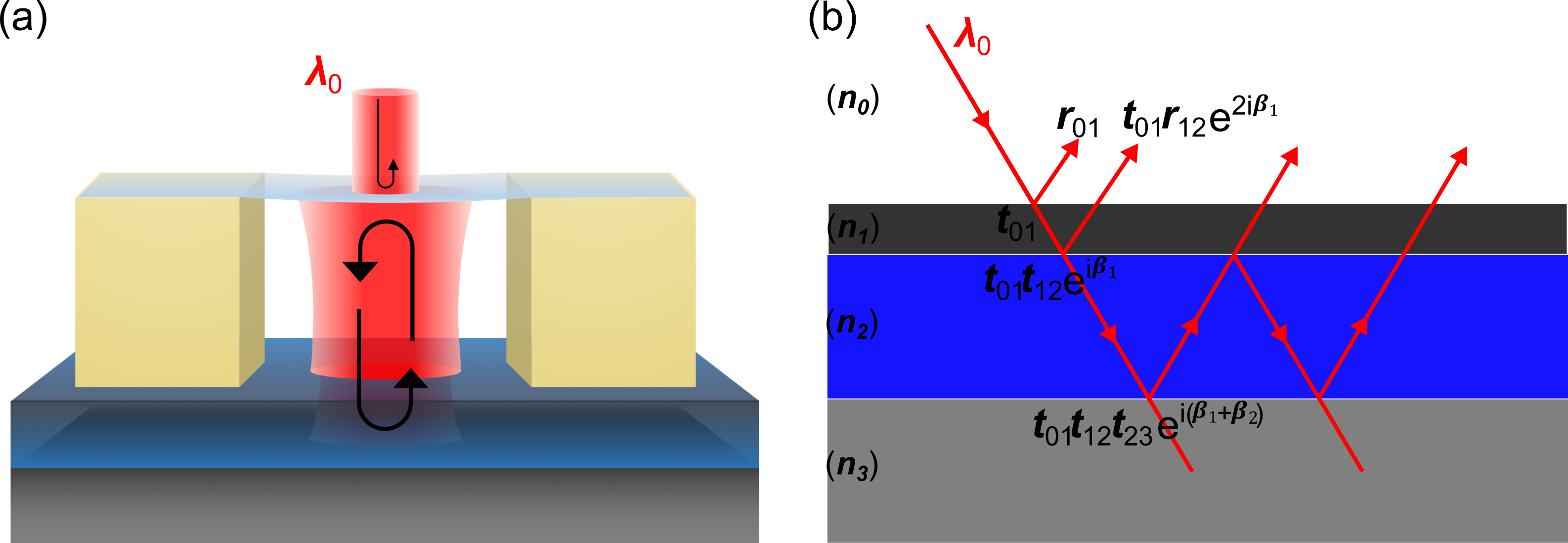}
	\caption[Schematic of absorption calculation]{Suspended-2DM optical cavity and optical interferences leading to absorbtion. (a) Schematic of the enhanced circulating power inside the optical cavity formed by a suspended BLG and substrate surfaces (mirrors). (b)Series of Fresnel reflections, $r_{ij}$, and transmissions, $t_{ij}$, at the various interfaces between media $i$ and $j$.}
	\label{fig:S11}
	\end{figure}

\Cref{fig:S11}a is a representation of the enhanced circulating power inside the optical cavity formed by a suspended BLG and back-plan substrate (Si) due to interferences. This can lead to increased light absorption when the mirror spacing creates constructive interferences. To describe this quantitatively, in \Cref{fig:S11}b we show the series of Fresnel reflections, $r_{ij}$, and transmissions, $t_{ij}$, at the various interfaces between media $i$ and $j$. The reflection (transmission) coefficient values are the ratio of the amplitude of the reflected (transmitted) light's electric field to the incident light. They are given numerically by $r_{ij} = (n_i-n_j)/(n_i+n_j)$ and $t_{ij} = (2n_i)/(n_i+n_j))$, where $n_{i}$ and $n_{j}$ are the complex indices of refraction of material, and they are wavelength dependent. The real part of the index of refraction conserves energy, while the imaginary part absorbs energy (dissipative currents). The term $\beta_i$ is the complex phase shift introduced while light travels in medium $i$, and depends on $n_i$, the medium's thickness $h_i$, and incident light wavelength $\lambda$, as per Equation \ref{eq:beta1}. \Cref{tab:S3} below summarizes the indices of refraction we used in our calculations.

\begin{equation}\label{eq:beta1}	
	\beta_i = \frac{ 2\pi n_i h_i} {\lambda}
	\end{equation}

	\strutlongstacks{T}
	\begin{table}[H]
	\begin{tabular}{ l c c }
 	\toprule[1pt]\midrule[0.3pt]
 	\multicolumn{1}{c}{Material} & Real part of $n$ & Imaginary part of $n$ \\ \midrule
	BLG (532 nm) \cite{Ghamsari16}	&	2.33	&	1.3	\\
	BLG (581 nm) \cite{Cheon14}	&	2.41	&	1.373	\\
	BLG (621 nm) \cite{Cheon14}  &	2.47	&	1.43	\\ \midrule[0.3pt]
	Al (532 nm)	\cite{Mcpeak15}	&	0.636	&	5.38	\\
	Al (581 nm) \cite{Mcpeak15}	&	0.794	&	5.88	\\
	Al (621 nm)	\cite{Mcpeak15}	&	0.948	&	6.26	\\ \midrule[0.3pt]
	\ch{SiO2} (532 nm) \cite{Gao12}	&	1.475	&	0	\\
	\ch{SiO2} (581 nm) \cite{Gao13}	&	1.473	&	0	\\
	\ch{SiO2} (621 nm)	\cite{Gao13}&	1.47		&	0	\\ \midrule[0.3pt]
	Si (532 nm) \cite{Green08}	&	4.14		&	0.033	\\
	Si (581 nm) \cite{Schinke15}	&	3.98		&	0.22		\\
	Si (621 nm)	\cite{Schinke15}	&	3.89		&	0.017	\\
 	\midrule[0.3pt]\bottomrule[1pt]
	\end{tabular}
	\caption{\label{tab:S3}List of all indices of refraction for different wavelength and material.}
	\end{table}

We first show how we calculate a $N$-layer heterostructure's total reflection coefficient, $r_{N}$, and transmission coefficient, $t_{N}$, starting from the individual interfacial coefficients $r_{ij}$ and $t_{ij}$. Then we show how $r_{N}$ and $t_{N}$ leads to an expression for $A_{BLG}$, the exclusive light absorption of BLG when inserted in a planar heterostructure.

For a system with $N=$ 3 media (layers), we find that \cite{Hecht02},
	\begin{subequations}\label{eq:rt3}
	\begin{align}
	r_{3} &= \frac{ r_{01} + r_{12} e^{2i\beta_1} } { 1+r_{01} r_{12} e^{2i\beta_1} }
	\label{eq:rt3a}	\\[10pt]
	t_{3} &= \frac{ t_{01} t_{12} e^{i\beta_1} } { 1+r_{01} r_{12} e^{2i\beta_1} }
	\label{eq:rt3b}
	\end{align}
	\end{subequations}

where $r_{01}, r_{12}, t_{01}$ and $t_{12}$ are the reflection and transmission coefficients (Fresnel’s) for light rays at the interfaces of media 0 and 1, and 1 and 2 respectively. We can simply use a recursive method to obtain all of the $r_{N}$ and $t_{N}$, when $N > 3$. The recursive expression to derive the $N$ coefficients from the $N-1$ ones, is given by Equation \ref{eq:recurrence}. Note that in Equation \ref{eq:recurrence}, the newly added medium is placed on top of the previous $N-1$ media (see \Cref{fig:S11}b), and is now assigned the label ``0'' while the other media’s labels are increased by 1.

	\begin{subequations}		\label{eq:recurrence}
	\begin{align}
	r_{N} &= \frac{ r_{01} + r_{N-1} e^{2i\beta_1} }
	{ 1+r_{01} r_{N-1} e^{2i\beta_1} }		\label{eq:recurrencea} 	\\[10pt]
	t_{N} &= \frac{ t_{01} t_{12} e^{i\beta_1} }
	{ 1+r_{01} r_{N-1} e^{2i\beta_1} } \ , 	\label{eq:recurrenceb}
	\end{align}
	\end{subequations}

Thus, in order to derive the coefficients for a $N=$ 4 media system, we simply plug the coefficients for $N=$ 3 from Equations \ref{eq:rt3a} and \ref{eq:rt3b} into Equations \ref{eq:recurrencea} and \ref{eq:recurrenceb}, and find:

	\begin{subequations}		\label{eq:rt4}
	\begin{align}
	r_{4} &= \frac{ r_{01} + r_{12} e^{2i\beta_1}
	+(r_{01} r_{12} + e^{2i\beta_1} ) r_{23} e^{2i\beta_2} }
	{ 1+r_{01} r_{12} e^{2i\beta_1}  +(r_{12} + r_{01} e^{2i\beta_1}) r_{23} e^{2i\beta_2} }
	\label{eq:rt4a} \\[10pt]
	t_{4} &= \frac{ t_{01} t_{12} t_{23} e^{i (\beta_1+\beta_2)} }
	{ 1+r_{01} r_{12} e^{2i\beta_1} +(r_{12} + r_{01} e^{2i\beta_1}) r_{23} e^{2i\beta_2} } \ . 	\label{eq:rt4b}
	\end{align}
	\end{subequations}

The fraction of power reflected, $R_{N}$, by a $N$ media heterostructure is given by squaring the complex amplitude, $R_{N} = \abs{r_{N}}^2$ . The back plane substrate is a semi-infinite (very thick) medium and there is no transmission across it\cite{Song15}. Thus,

	\beq \label{eq:AN}
	A_{N}= 1 - R_{N}
	\eeq

Our objective is to calculate specifically (exclusively) the light absorbed by the BLG (or another specific 2DM replacing it). In the heterostructures (optical cavities) we study, the 2D layers used as spacers between the top BLG and bottom mirror (substrate) are air, hBN, or SiO$_2$. They all have purely real indices of refractions, and do not lead to any absorption. The only material dissipating power, in addition to the BLG, is the bottom substrate/mirror, made of either Si or Al. We therefore, calculate the exclusive light absorption of BLG as:

	\beq \label{eq:ABLG}
	A_{\mathrm{BLG}}=1-R_{N}-A_{substrate}
	\eeq

\noindent where $A_{\mathrm{substrate}} = A_{\mathrm{Si}}$ in our BLG/air/SiO$_2$/Si devices, and $A_{\mathrm{substrate}} = A_{\mathrm{Al}}$ in our BLG/hBN/Al devices. Based on \Cref{fig:S11}b, we can identify $A_{\mathrm{substrate}} = T_{N} = \frac{n_N}{n_0} \abs{t_{N}}^2$, which can readily be calculated using Equation \ref{eq:recurrenceb}. We mention that the wavelength used in Equation \ref{eq:ABLG} is the incident laser wavelength.

\textit{Raman factors in BLG} -- While it is simple to calculate $A_{\mathrm{BLG}}$, experimentally it is rather challenging to isolate it from other absorption processes. To achieve an optical measurement which contains a unique BLG fingerprint, we use Raman spectroscopy. While we cannot measure directly light absorption with Raman, we can measure closely related quantities called the Raman Factors, $F_{\mathrm{BLG-G}}$ for the G-peak and $F_{\mathrm{BLG-2D}}$ for the 2D-peak. The Raman factor's meaning is the ration between the Raman intensity observed in a material (integrated Raman count) inside an heterostructure (i.e. including interferences) and the Raman intensity in the same material surrounding by only vacuum. Thus experimentally, $F_{\mathrm{BLG}}$ = (BLG-in-heterostructure Raman count) / (BLG-in-vacuum Raman count).

The $F_{\mathrm{BLG}}$'s are calculated with the same Fresnel coefficients as $A_{\mathrm{BLG}}$, as described below. Thus their experimental measurement can confirm that values necessary to extract the BLG absorption, in addition to demonstrating the tunability of the Raman scattering intensities. Moreover, the ability to measure \textit{simultaneously} both of the Raman Factors (same sample, same time, same laser, same systematic errors) in several devices, provides a very robust experimental verification of the calculations.

	\begin{figure}
	\includegraphics[scale=1]{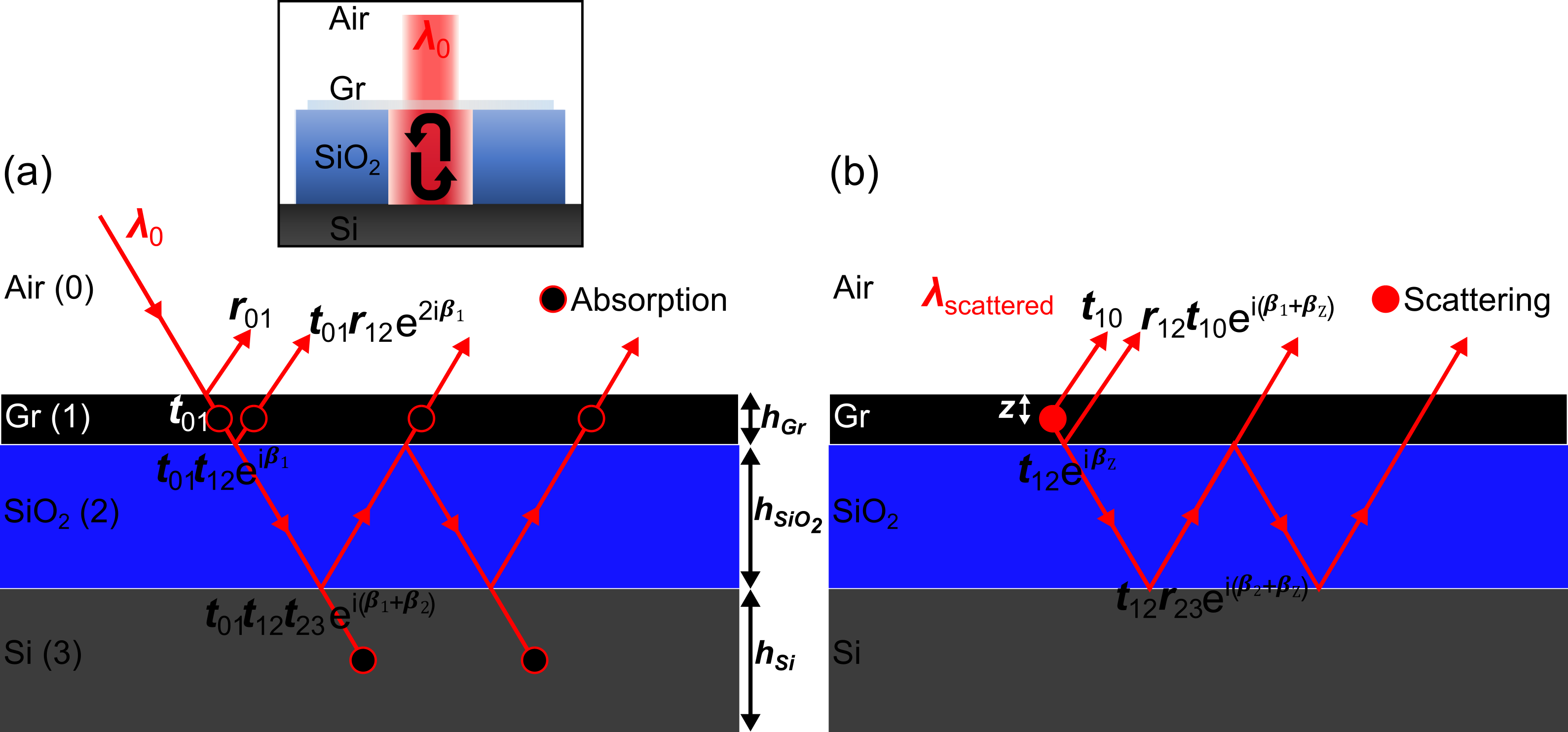}
	\caption[Schematic of Raman reflection interference]{Schematic of reflection interferences both before (a) and after (b) Raman scattering. (a) $\bm{E}_0$ is the incident laser electric field, $\bm{E}_z$ is the field inside medium 1. The Fresnel equations are used to calculate the transmission and reflection amplitudes. (b) We start with the $\bm{E}_{\mathrm{Ram}}$, Raman scattered E-field produced inside medium 1 (BLG), to calculate the reflections and transmissions coefficients following the scattering event. This Raman-shifted light also goes through multiple reflections and transmissions before exiting as $\bm{E}_{\mathrm{out}}$.}
	\label{fig:S12}
	\end{figure}

To calculate the $F_{\mathrm{BLG}}$'s, we must first calculate the absorption amplitude for $N$-media structure, $F_{ab-N}$, from the interferences due to multiple reflection/transmission paths at the incident laser light (see \Cref{fig:S12}a) as well the amplitudes, $F_{sc-N-G}$ or $F_{sc-N-2D}$, due to the multiple reflection/transmission paths of the Raman-shifted light after scattering, as shown in \Cref{fig:S12}b \cite{Wang08}. We follow \textit{Yoon et al.}\cite{Yoon09}, and find for a $N =$ 3 media system,

	\begin{subequations}		\label{eq:F3}
	\begin{align}
	F_{ab\mathrm{-3}}(\lambda = \lambda_{laser}) &=
	t_{01} \frac{ e^{-i \beta_z} + r_{12} e^{-i (2\beta_1-\beta_z)} }
	{ 1+ r_{12} r_{01} e^{-2 i \beta_1} } 	\label{eq:F3a}	
	\\[10pt]
	F_{sc\mathrm{-3}}(\lambda = \lambda_{Raman-shifted}) &=
	t_{10} \frac{ e^{-i \beta_z} + r_{12} e^{-i (2\beta_1-\beta_z)} }
	{ 1+ r_{12} r_{01} e^{-2 i \beta_1} }	\ , 	\label{eq:F3b}
	\end{align}
	\end{subequations}

where $\beta_z$ is the complex phase shift introduced along the BLG thickness at position $z$. It is important to notice that Equation \ref{eq:F3b} includes $t_{10}=2n_1/(n_1+n_0) \neq t_{01}$, and that the wavelength used is the one after the Raman shift, while the original (laser) wavelength is used in Equation \ref{eq:F3a}. Our incident laser has a wavelength of 532 nm, the shifted laser wavelength for the G peak is 581 nm (1580 cm$^{-1}$), and for the 2D peak it is 621 nm (2700 cm$^{-1}$). Equations \ref{eq:F3} can also be extended to planar heterostructures made of $N$-media as follows,
	
\begin{subequations}		\label{eq:FN}
	\begin{align}
	F_{ab\mathrm{-N}}(\lambda = \lambda_{laser}) &=
	t_{01} \frac{ e^{-i \beta_z} + r_{\mathrm{N-1}} e^{-i (2\beta_1-\beta_z)} }
	{ 1+ r_{\mathrm{N-1}} r_{01} e^{-2 i \beta_1} } \label{eq:FNa}	
	\\[10pt]
	F_{sc\mathrm{-N}}(\lambda = \lambda_{Raman-shifted}) &=
	t_{10} \frac{ e^{-i \beta_z} + r_{\mathrm{N-1}} e^{-i (2\beta_1-\beta_z)} }
	{ 1+ r_{\mathrm{N-1}} r_{01} e^{-2 i \beta_1} }	\ . 		\label{eq:FNb}
	\end{align}
	\end{subequations}

For instance, when $N=4$ we find,
	
    \begin{subequations}		\label{eq:F4}
	\begin{align}
	F_{ab\mathrm{-4}}(\lambda = \lambda_{laser}) &=
	t_{01} \frac{ (1+r_{12} r_{23} e^{-2i\beta_2}) e^{-i \beta_z} + (r_{12}+r_{23} e^{-2i\beta_2} ) e^{-i (2\beta_1-\beta_z)} }
	{ 1+ r_{12} r_{23} e^{-2i\beta_2}+(r_{12} + r_{23} e^{-2i\beta_2}) r_{01} e^{-2 i \beta_1} }
	\label{eq:F4a} \\[10pt]
	F_{sc\mathrm{-4}}(\lambda = \lambda_{Raman-shifted}) &=
	t_{10} \frac{ (1+r_{12} r_{23} e^{-2i\beta_2}) e^{-i \beta_z} + (r_{12}+r_{23} e^{-2i\beta_2} ) e^{-i (2\beta_1-\beta_z)} }
	{ 1+ r_{12} r_{23} e^{-2i\beta_2}+(r_{12} + r_{23} e^{-2i\beta_2}) r_{01} e^{-2 i \beta_1} } \ . \label{eq:F4b}
	\end{align}
	\end{subequations}

Finally, to calculate the desired Raman Factors, $F_{BLG-G}$ and $F_{BLG-2D}$, we use the above amplitudes following \textit{Yoon et al.}\cite{Yoon09},

	\begin{subequations}		\label{eq:FBLG}
	\begin{align}
	F_{BLG-G}=\mathcal{N}_{G} \int_0^{h_{BLG}} \abs{F_{ab}\cdot F_{sc-G} }^2 dz	\label{eq:FBLGa} 	\\[10pt]
	F_{BLG-2D}=\mathcal{N}_{2D} \int_0^{h_{BLG}} \abs{F_{ab}\cdot F_{sc-2D} }^2 dz \ , 	\label{eq:FBLGb}
	\end{align}
	\end{subequations}

\noindent and $\mathcal{N}$ is a normalization constant (1 divided by the integral calculated for BLG surrounded by vacuum only). The integration is over the thickness of the BLG crystal, where z is a dummy variable representing the vertical position inside the graphene, see \Cref{fig:S12}.

	\begin{figure}
	\includegraphics[scale=1]{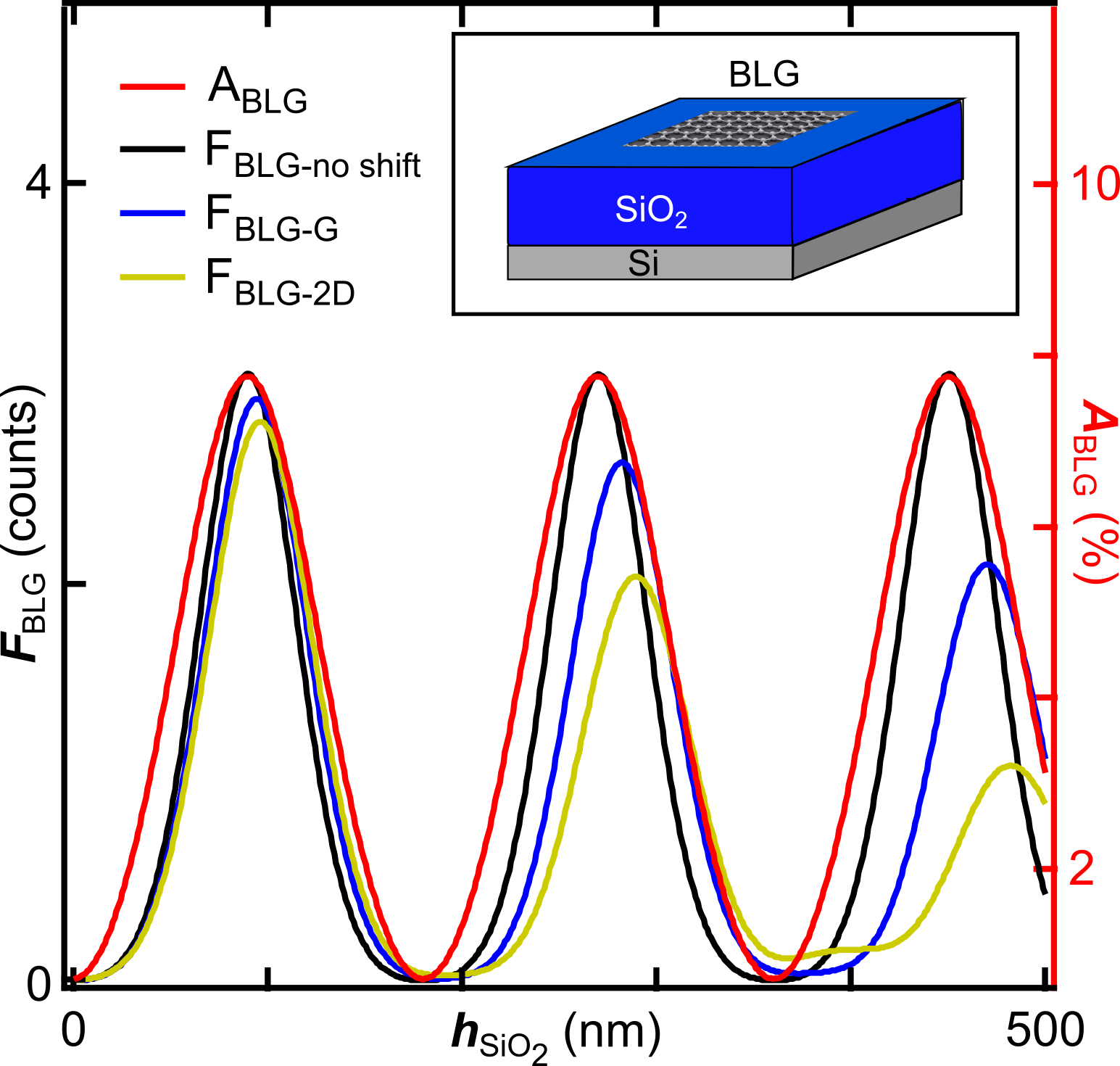}
	\caption[Raman factor vs wavelength]{Raman Factors and exclusive light absorption of BLG in Air/BLG/\ch{SiO2}/Si heterostructure. The left axis shows $F_{\mathrm{BLG-no shift}}$ (black), $F_{\mathrm{BLG-G}}$ (blue), and $F_{\mathrm{BLG-2D}}$ (gold) versus $h_{\mathrm{SiO_{2}}}$, and the right axis display $A_{\mathrm{BLG}}$. The inset shows a diagram of the heterostructure geometry. }
	\label{fig:S13}
	\end{figure}

\Cref{fig:S13} shows Raman factors and absorption calculated in BLG/SiO$_2$/Si heterostructures (inset) as a function of the thickness of the SiO$_2$ film. The solid black trace shows $F_{\mathrm{BLG-no shift}}$, is the Raman Factor calculated under the approximation that there is zero Raman wavelength shift. This is to be compared with the $F_{\mathrm{BLG-G}}$ (blue trace) and the $F_{\mathrm{BLG-2D}}$, which clarify the impact of the wavelength shift on the predicted Raman intensity. Lastly, the red trace shows the calculated $A_{\mathrm{BLG}}$ (right axis), whose maxima align well with $F_{\mathrm{BLG-no shift}}$.

As an example, for our BLG/310$\pm$3nm-SiO$_2$/Si heterostructures, we can read from \Cref{fig:S13} the calculated $F_{\mathrm{BLG-G}}=$ 1.48 $\pm$ 0.18 and $F_{\mathrm{BLG-2D}}=$ 1.53 $\pm$ 0.12, where the errors are systematic ones (i.e. a rigid shift of the whole data sets depending on the exact SiO$_2$ thickness on the wafers). The corresponding exclusive BLG light absorption in \Cref{fig:S13} is 4.25 \%. We emphasize that these calculations do not include any fitting or free parameter.

\subsection{Experimental Raman intensity measurements, and calibration of the Raman factors}\label{sec:S4.2}

To measure experimentally Raman factors and make a quantitative comparison with the calculated ones, we first explain how to extract the raw experimental Raman counts, and then we discuss how to convert these Raman counts into Raman factors based on reliable calibration devices.

\begin{figure}
	\includegraphics[scale=1]{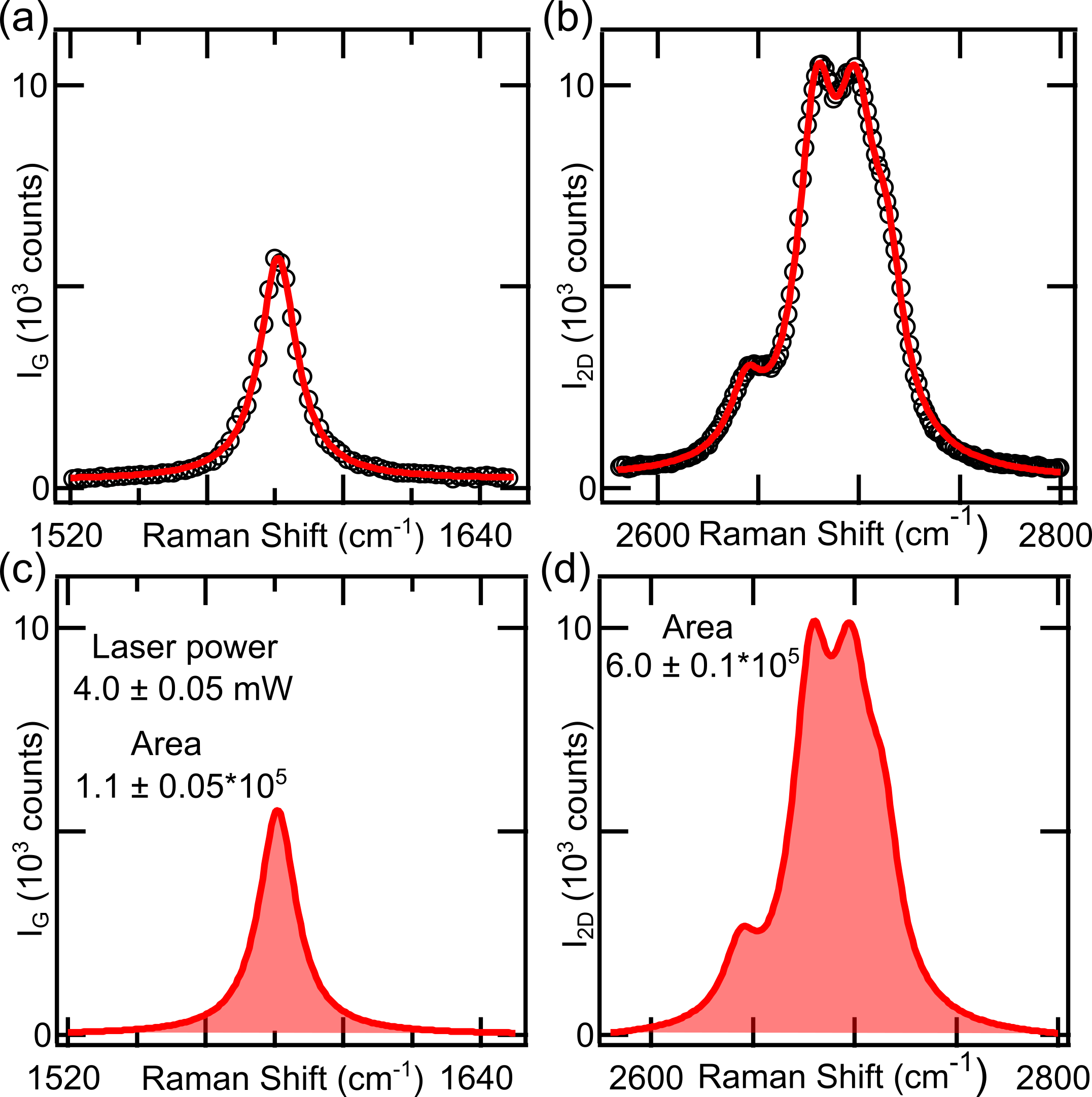}
	\caption[Curve fitting]{Curve fitting to extract the integrated Raman intensities in a BLG/310nm-SiO$_2$/Si device. (a) Fitting the G-peak and (b) the 2D-peak. The open black circles are the raw data and the red traces are the fit functions for G and 2D peak. (c) Integrated Raman count under the G-peak fit, and (d) under the 2D-peak fit.}
	\label{fig:S14}
	\end{figure}

\textit{Experimental Raman intensity measurements} -- The Raman counts for the G and 2D peaks in BLG are measured by integrating the areas under the measured peaks. In order to extract these numbers, we first do a fit of the experimental curves using a single Lorentzian (G-peak) or a combination of four Lorentzian functions (2D-peak). Examples of these fits are shown respectively in \Cref{fig:S14}a-b for data from a BLG/310nm-SiO$_2$/Si device. Using these fits, we remove the background signal (vertical offset and linear background) from the raw data, and then integrate the area under the peaks as shown in \Cref{fig:S14}c-d. The integrated counts shown in \Cref{fig:S14}c-d are typical and have very low uncertainties, and scale linearly with both laser power and integration time. The range of integration is set such that, increasing its range does not change the extracted number of counts.

\begin{figure}
	\includegraphics[scale=1]{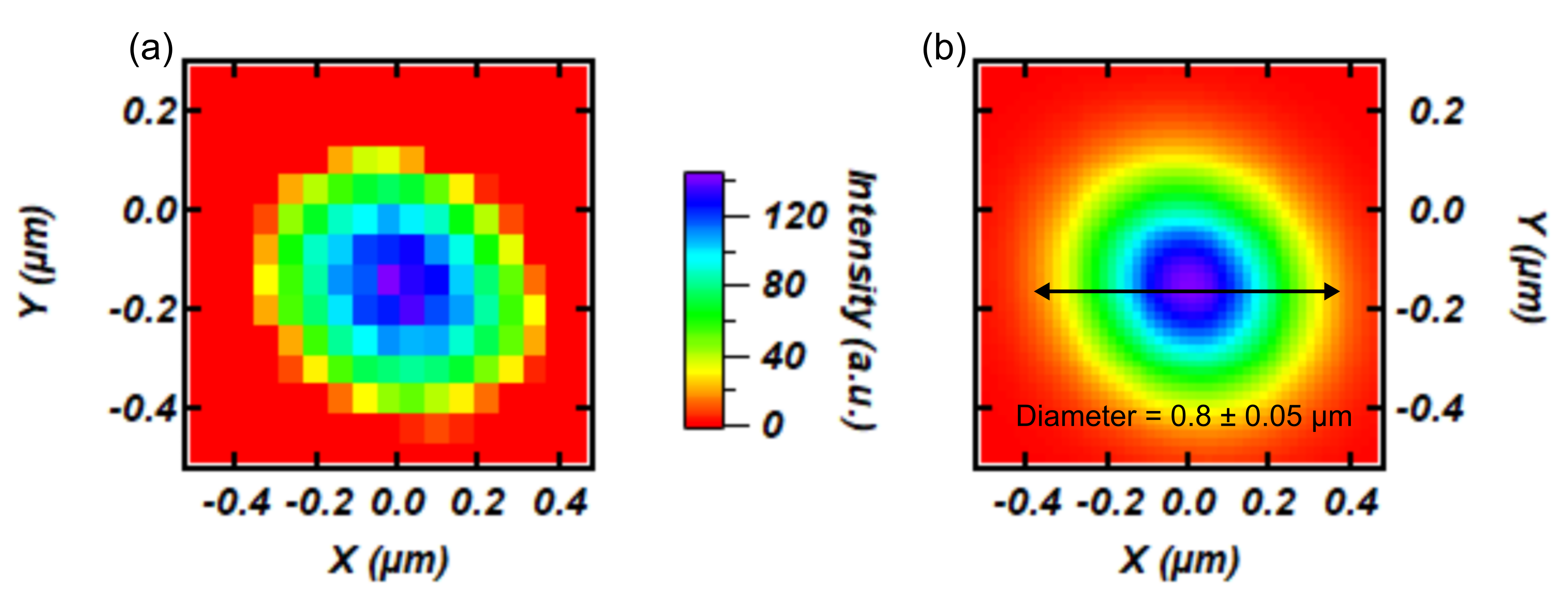}
	\caption[Laser spot]{Laser beam intensity profile. (a) The measured intensity profile of our focused laser beam at 532 nm. (b) The fitted laser intensity profile. It gives beam width of 0.8 $\mu$m, which sets the spatial resolution of our Raman measurements.}
	\label{fig:S15}
	\end{figure}

\Cref{fig:S15}a show the measured, and \Cref{fig:S15}b the fitted, laser intensity profile of our laser beam. The width of our focused laser is 0.8$\pm$0.05 $\mu$m and determines our spatial resolution.

\textit{Calibrating the experimental Raman factors} -- The theoretical calculations of $F_{\mathrm{BLG}}$ discussed above do not involve any fitting parameter, but we need one single reference calibration (conversion factor) to relate the many measured Raman counts to the calculated Raman Factors. For this calibration, we used a series of 5 identically prepared calibration samples air/BLG/310 nm-\ch{SiO2}/Si. Using the above calculated Raman Factors for such heterostructures ($F_{\mathrm{BLG-G}}=$ 1.48 and $F_{\mathrm{BLG-2D}}=$ 1.53), and the measured Raman counts in Figure \Cref{fig:S14} we can establish a calibration factor relating experiment and theory.

\begin{figure}
	\includegraphics[scale=1]{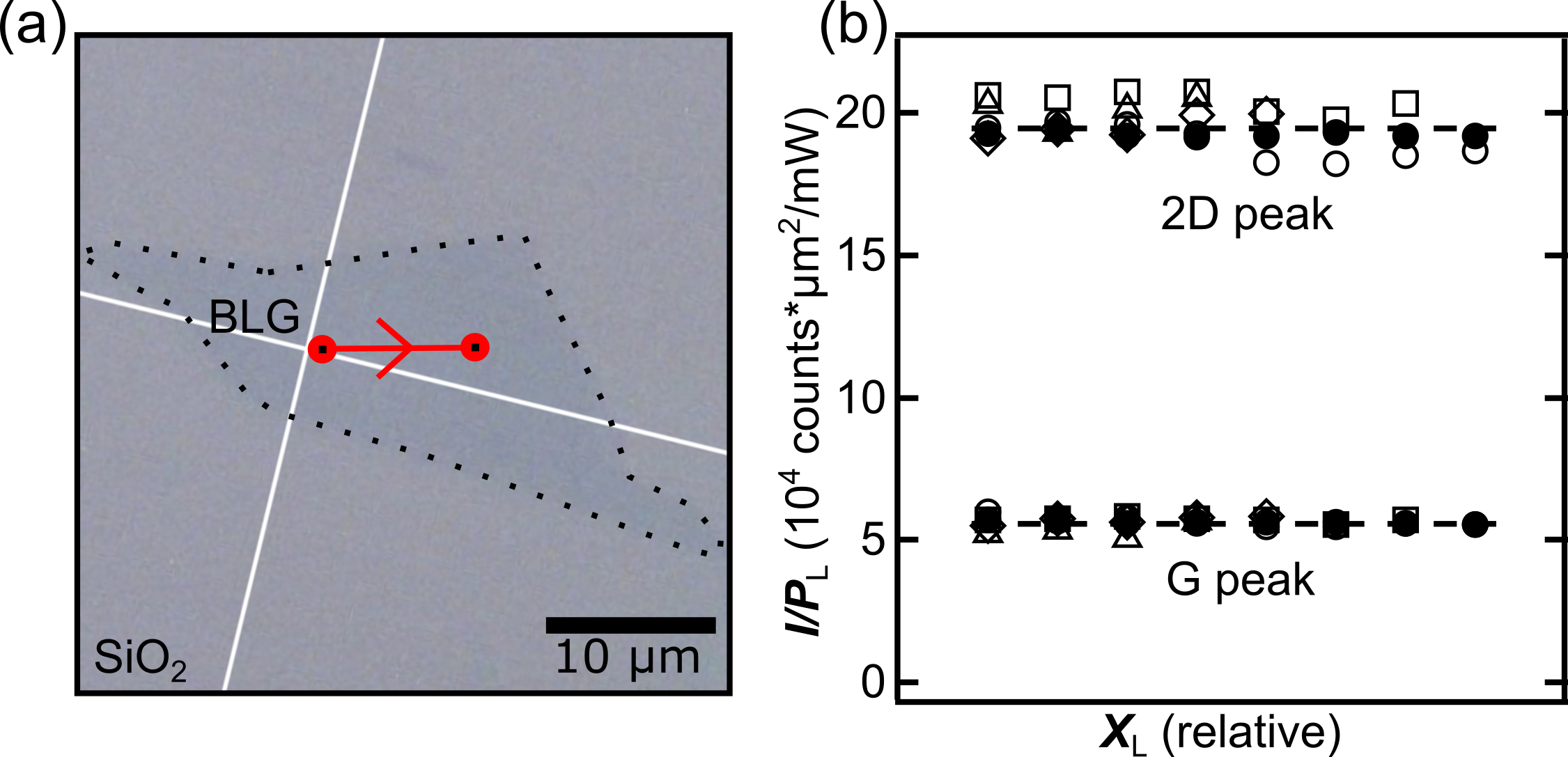}
	\caption[Raman factor calibration]{Calibration of the integrated Raman counts to the calculated Raman factors. (a) Top view of a calibration BLG/310nm-SiO$_2$/Si heterostructure. The red line shows the locations of the Raman data acquisitions. (b) The normalized integrated Raman counts $I/P_{L}$, for 10s exposures, measured in five devices similar to the one in (a) at many different positions $X_{L}$, for both the G-peak and 2D-peak.}
	\label{fig:S16}
	\end{figure}

The details of this calibration are shown in \Cref{fig:S16}. In panel (a) we show one of the five air/BLG/310 nm-\ch{SiO2}/Si reference samples and indicate in red the path along which many independent Raman measurements were made. The normalized Raman counts $I/P_{L}$, from dividing the integrated counts by the (laser power/area $\times$ exposure time) used during the acquisitions, is shown in \Cref{fig:S16}b versus the position along the red line shown in \Cref{fig:S16}(a). Each set of symbols in (b) represents the data from one for the five devices. The laser powers and acquisition times used were 1 – 5 mW/$\mu\mathrm{m}^2$, and 10 s respectively. We found a very high consistency in the measured $I/P_{L}$ in different samples and at different locations, for both the G-peak and 2D-peak. We used the average values of $I/P_{L}$ for the G and 2D resonances, and assign them the calculated Raman factors. This gives us a the following conversion constant of $F_{\mathrm{BLG-G}}=$ 1.0 $=$ 38000 $\pm$1000, and $F_{\mathrm{BLG-2D}}=$ 125000 $\pm$ 4000 counts for a laser power of 1 ± 0.02 mW/$\mu\mathrm{m}^2$ and 10 second of exposure. We verified carefully that the calibration ratio is unchanged by the experimental laser power in a range well beyond the values used in our data acquisition (1 – 20 mW/$\mu\mathrm{m}^2$). We used this calibration for all reported experimental $F_{\mathrm{BLG}}$'s.

\textit{Error bars in Figure 4 of the main text}
The vertical uncertainties reported for the experimental data in Figure 4b-c, e-f, and h-i, stem primarily from the uncertainty of $\pm$ 3 nm on the thickness of the SiO$_2$ film on our Raman factor calibration devices. As explained in the paragraphs above, a change in SiO$_2$ thickness leads to a change in the calculated Raman factors, and thus in the conversion (calibration) constant for the experimental $F_{\mathrm{BLG-G}}$ and $F_{\mathrm{BLG-2D}}$. The horizontal uncertainties in the data reported in Figure 4 stems from the uncertainties on thickness of the optical cavity spacers ($h_{hBN}$ or $h_{air}$. This uncertainty itself stems from various sources including, the thickness variations of the hBN crystals or suspension heights at various laser positions, the precision of laser beam positions (around $\pm$ 0.5 $\mu$m), AFM  data noise, and AFM calibration limitations. Specifically for Figures 4h-i, which refer to data from a tilted-suspended device, an additional source of horizontal-axis uncertainty is the cross-section profile (slope) of the suspended device. Based on both the optical images and Raman data, we found that the suspension slope $dz/dx$ was roughly linear. To establish a realistic uncertainty for this approximation, we compared (took the difference between) the assumed linear slope positions and the AFM measured positions in a similar tilted-suspended device. This explains why the horizontal error bars in Figures 4h-i vary along the x-axis.

\clearpage
\BeginNoToc
\bibliography{Rebollo_Rodrigues_SI_submitted}
\EndNoToc